\documentclass[lettersize,journal]{IEEEtran}
\usepackage{amsmath,amsfonts}

\usepackage{cite}
\usepackage{amsmath,amssymb,amsfonts}
\usepackage{graphicx}
\usepackage{textcomp}
\usepackage{array}
\usepackage{multirow}
\usepackage{graphicx}
\usepackage{subfigure}
\usepackage{cases}
\usepackage{algpseudocode}
\usepackage{algorithm}
\usepackage{color}
\usepackage{soul}
\usepackage{xcolor}
\usepackage[switch]{lineno}
\usepackage{tabularray}

\begin{document}

\title{Fuzzy Q-learning-based Opportunistic Communication for MEC-enhanced Vehicular Crowdsensing}

\author{Trung Thanh Nguyen\IEEEauthorrefmark{2}, Truong Thao Nguyen\IEEEauthorrefmark{3}, Thanh-Hung Nguyen\IEEEauthorrefmark{2}\IEEEauthorrefmark{1}, Phi Le Nguyen\IEEEauthorrefmark{2}\IEEEauthorrefmark{1}

\IEEEcompsocitemizethanks{
\IEEEcompsocthanksitem \IEEEauthorrefmark{2}School of Information and Communication Technology, Hanoi University of Science and Technology, Hanoi, Vietnam. 
E-mail: \{thanh.nt176874@sis, hungnt@soict, lenp@soict\}.hust.edu.vn
\IEEEcompsocthanksitem \IEEEauthorrefmark{3}The National Institute of Advanced Industrial Science and Technology (AIST), Japan. E-mail: nguyen.truong@aist.go.jp
\IEEEcompsocthanksitem \IEEEauthorrefmark{1}Corresponding authors.
}
}



\maketitle

\begin{abstract}
This study focuses on MEC-enhanced, vehicle-based crowdsensing systems that rely on devices installed on automobiles. We investigate an opportunistic communication paradigm in which devices can transmit measured data directly to a crowdsensing server over a 4G communication channel or to nearby devices or so-called Road Side Units positioned along the road via Wi-Fi. 
We tackle a new problem that is how to reduce the cost of 4G while preserving the latency. 
We propose an offloading strategy that combines a reinforcement learning technique known as Q-learning with Fuzzy logic to accomplish the purpose.
Q-learning assists devices in learning to decide the communication channel. Meanwhile, Fuzzy logic is used to optimize the reward function in Q-learning.
The experiment results show that our offloading method significantly cuts down around 30-40\% of the 4G communication cost while keeping the latency of 99\% packets below the required threshold.
\end{abstract}


\section{Introduction}\label{sec:introduction}
With substantial advances in sensing, communication, and mobile computing technologies in recent years, a new computing and sensing paradigm named mobile crowdsensing has demonstrated an effective solution for the capillary gathering of huge quantities of information in densely populated regions \cite{3185504}.
In a crowdsensing system, participants equipped with sensing and computing capabilities work collaboratively to collect, share, and extract data about a phenomenon of common interest.
In the past, crowdsensing systems have mainly relied on mobile devices such as smartphones. 
Nowadays, vehicles with increasing sensing, computing, and storage capabilities have emerged as viable alternatives for mobile crowdsensing systems.
Numerous vehicle-based crowdsensing applications have been proposed, including traffic monitoring and prediction, advertisement dissemination \cite{7517783, 7192632, 6193237}. 
The conventional crowdsensing paradigm is a centralized cloud-based method in which data is sent from participants to a cloud server over a broadband network \cite{8703108}.
This approach, however, generates significant traffic on the network and computation burden on the cloud; thus, it cannot efficiently support real-time and large-scale mobile crowdsensing systems. 
To this end, one solution is to deploy Mobile Edge Computing (MEC) servers to Roadside Units (RSUs) that are close to vehicles. With the advantage of the closeness to the vehicles, MEC can help collect data from vehicles quickly and reduce the load on the radio access network.
In the literature, significant efforts have been devoted to MEC-enhanced vehicle-based mobile crowdsensing systems with various topics. 
Authors in \cite{9260141} focused on participant selection and task offloading problems. Liu et al. in \cite{9162762} leveraged meta-heuristic to address the network selection and traffic allocation. 
The authors in \cite{8662620} proposed a hierarchical task allocation framework which consists of two tiers: cloud and edge. The cloud layer evaluates participants' reputations and offers the most promising candidates to the edge layer. The edge layer then contacts the participants and optimizes the task allocation. In \cite{Xia2019}, the authors introduced a quality-aware sparse data collecting technique. The goal is to guarantee spatiotemporal coverage while minimizing redundant data. The main idea is to leverage the correlation among sensing data to identify the smallest subset of grids to allocate tasks. Using correctly acquired data, the cloud server then infers the missing values.
Zhao in \cite{Zhao2021} designed an optimal sensing strategy for all vehicles. 

Different from existing works, this study considers a novel problem that asks to minimize the communication budget while maintaining the freshness of information in MEC-enhanced Vehicle-based Mobile Crowdsensing systems (MVMC).
Specifically, we focus on MVMC systems, where the sensory data from vehicles are opportunistically transferred to the server via three communication routes: (1) directly sending to the server through the cellular networks such as 4G, (2) transferring to a roadside unit (RSU) via Wi-Fi and then transmitting from RSU to the server by the wired network, and (3) relaying to a nearby vehicle by using Wi-Fi, and then following the nearby vehicle's policy to transfer to the server.
The network model is illustrated in Fig.~\ref{fig:network_model}.
Our goal is to ensure data freshness while reducing transmission expenses.
We define data freshness as the amount of time between when the data was generated and when the cloud server collected it.
We then aim to lessen the fraction of packages with freshness levels surpassing a specified threshold. 
Furthermore, we assume that the 4G network is widespread and that the device can always send data to the cloud in real-time through 4G. 
On the other hand, Wi-Fi has a limited coverage area, so the device cannot always communicate data to the RSU or other devices.
The drawback of 4G over Wi-Fi is that the cost per communicated capacity on a 4G network is substantially higher (see Table \ref{tab:4Gcost}).
As a result, we should design an offloading mechanism that minimizes the number of packets sent directly from the device to the server (to save expenses) while maintaining packet freshness.
Additionally, 4G transmission uses much more energy than Wi-Fi.
For instance, we performed an experiment to determine how much energy an air quality monitoring device consumes when communicating through Wi-Fi or 4G.
The experimental findings in Table \ref{tab:energyCost} indicate that employing 4G consumes $1.3$ times of energy compared to Wi-Fi.
We name our targeted problem as OCVC (stands for Opportunistic Communication for Vehicle-based mobile Crowdsensing).
Our OCVC problem asks to minimize the use of 4G communication while guaranteeing that the information latency does not exceed a threshold. 
Here, the term \emph{information latency} is defined by the time interval from when the data is collected until it reaches the server. 

\begin{table}[t]
    \centering
    \caption{Average prices of 4G and Wired network communication over the world } \label{tab:4Gcost}
    \begin{tabular}{|c|c|c|c|}
    \hline
         \multirow{3}*{\emph{Regional}} & \multirow{1}*{\emph{4G communication }} & \emph{{Wired network }}\\
         & \emph{cost (\$)}  & \emph{communication cost (\$)} \\
         & \emph{(per 1GB data)} & \emph{(per 1 month)} \\
         \hline
         Global & 4.07 &  57.07 \\
         Asia & 1.79 & 40.29 \\
         Baltics & 2.09  & 19.19\\
         Caribbean & 4.44 & 78.44\\
         Central America & 2.40 & 43.87\\
         CIS (Former Ussr) & 2.84  & 13.96\\
         Eastern Europe & 4.64 & 19.90\\
         Near East & 3.94 & 60.62\\
         Northern Africa & 1.53 & 22.41\\
         Northern America & 8.21  & 89.44\\
         Oceania & 5.51 & 85.14\\
         South America & 5.52 & 55.17\\
         Sub-Saharan Africa & 6.44 & 77.70\\
         Western Europe & 2.47 & 49.56\\
         \hline
    \end{tabular}
\end{table}

To the best of our knowledge, this study is an early attempt to minimize the communication budget while maintaining the freshness of information in vehicular mobile crowdsensing.
Our idea is to exploit Q-learning combined with Fuzzy logic. 
In our Q-learning paradigm, each crowdsensing device is considered an agent that keeps track of its Q table.
We split the whole timeline into discrete time units called time slots.
Packets generated by a device are temporarily stored in the device's buffer.
The agent then uses the proposed Q-learning model to perform one of the following actions at every time slot: 1) storing data in the local buffer, 2) transferring to the RSU, 3) relaying to a neighbor device, and 4) transmitting to the server.
Each Q table item has a Q value that reflects how well an action performs in a particular state.
At each time slot, the agent will choose an action based on the value of the Q value (usually, actions with a higher Q value have a higher chance of being selected).
When the agent performs an action, the environment provides feedback indicating how beneficial the action was. This goodness is quantified by a so-called reward, which will be used to update the Q table.
Q-learning actions will be selected to maximize the cumulative reward value. Consequently, our reward function will be designed to encourage actions that help reduce 4G communication costs while preserving packet freshness.
Furthermore, we will exploit Fuzzy logic to adapt several hyperparameters in the reward function, making it more resistant to environmental changes.
Our contribution is as follows:
\begin{itemize}
    \item We formulate the opportunistic communication in MEC-enhanced vehicle-based mobile crowdsensing systems. 
    \item We employ Q learning to propose an offloading strategy that reduces communication costs while maintaining data freshness. Furthermore, to enhance the efficiency of the Q-learning-based offloading algorithm, we adopt Fuzzy logic to adjust Q-learning hyper-parameters adaptively.
    \item We perform extensive experiments to evaluate the impacts of various network configurations on the performance of the proposed algorithm and compare it with several baselines. The numerical results show that our proposed protocol outperforms the baselines.
\end{itemize}

The remainder of the paper is organized as follows. 
We briefly introduce the related works, and an overview of Q-learning and Fuzzy logic in Section ~\ref{sec:related_work} and \ref{sec:preliminaries}.
Sections \ref{sec:fuzzy_q_learning} describes our proposed protocol in details. 
We present the numerical results in Section \ref{sec:experimental} and conclude the paper in Section \ref{sec:conclusion_and_future_work}.

\begin{table}
    \centering
    \caption{Energy consumption of an air quality monitoring device consisting of sensors measuring PM2.5, PM10, NO2, CO2, SO2, humidity, temperature} \label{tab:energyCost}
    \begin{tabular}{|c|c|c|c|}
    \hline
         \emph{Parameters} & \emph{4G} & \emph{WiFi} \\
         \hline
         Supply voltage (V) & 7.4 & 7.4 \\
         Average current consumption (mA) & 305.68 & 236.56 \\
         Max current consumption (mA) & 707.42 & 441.54 \\
         Average power consumption (W) & 2.26 & 1.75 \\
         Max power consumption (W) & 5.24 & 3.28 \\
         \hline
    \end{tabular}
\end{table}

\section{Related work}\label{sec:related_work}
The OCVC problem can be categorized as an offloading problem in V2X (i.e., Vehicle-to-Everything) networks. 
As such, we begin with an overview of current research addressing various aspects concerning MEC-enhanced mobile crowdsensing applications.
Following that, we review the literature on MEC's offloading issue.

In \cite{9260141}, the authors define Local Edge Nodes (LENs) and Main Edge Node (MEN) that are responsible for selecting workers available in the area of interest. They then proposed an offloading mechanism to offload sensory data from the workers to the identified LENs. Liu et al. In \cite{9162762} leveraged meta-heuristic to address the network selection and traffic allocation problem. The objective is to maximize the users' transmission capability while minimizing the transmission delay. The authors first presented the mathematical formulation and then applied the PSO to provide a sub-optimal solution. 
The authors in \cite{8662620} proposed a hierarchical task allocation framework. Firstly, the cloud layer evaluates the participants’ reputations based on various data and sends the optimal subset of participants to the edge layer. The edge layer then interacts with the participants and performs task-specific optimizations.
In \cite{Xia2019}, the authors presented a data collecting method aimed at minimizing data redundancy while maintaining sensor grids' spatiotemporal coverage. To do this, the authors used correlations between sensing data to determine which user group should be selected. Additionally, the compressive sensing technique was utilized to retrieve data from the whole sensing region. In \cite{8924663} the authors proposed a novel mobile crowdsensing paradigm where mobile users can act as mobile MEC nodes. The authors designed a probabilistic model and an algorithm to select appropriate users for acting as mobile MEC nodes.
The authors in \cite{9501007} provided a crowd-sensing-assisted vehicular distributed computing mechanism to update a High-definition map (HD Map) for autonomous driving. In addition, the authors proposed a heuristic best-effort algorithm for crowdsensing nodes selection and tasks allocation with the goal of minimizing communication load.
Xu et al. in \cite{9527332}  investigated the data uploading problem in mobile crowdsensing systems. They proposed a mechanism for multiple edge nodes to collaborate to match a team of edge nodes with a sensing worker to satisfy the demand. The authors first proved the NP-hardness of the targeted problem and then introduced an algorithm based on Lagrangian relaxation. In \cite{9162762}, the authors addressed the cost reduction issue in MEC-enhanced vehicular crowdsensing systems with various data kinds carried by users. They proposed a mechanism for determining which server should be enabled for each data type of processing.
The authors in \cite{8648047} focused on collaborative mobile crowdsensing systems, where users are divided into groups and collaboratively exchange information inside each group. 
For each group, there is one owner that is responsible for gathering data from other members and forwarding it to the collector. 
The authors then proposed three grouping algorithms: static grouping, PoI grouping, and Dynamic grouping.
The experimental results show that users can save a large amount of energy by using the proposed grouping methods. Gong et al., in \cite{9013325}, studied the data offloading issue in opportunistic social networks, wherein mobile users may decide to offload data using their cellular interface or relay it to surrounding users to reduce costs.
The authors formulated the targeted problem mathematically and then proposed online and offline solutions.
The numerical findings showed that taking advantage of opportunistic offloading between users may significantly reduce costs.
K. Zhang et al. concentrated on optimizing energy usage in MEC-enabled 5G networks in \cite{7553459}.
The authors provide a mathematical model for the problem and propose an approximation approach.
The work in \cite{platoon_1, platoon_3, platoon_2} addressed the offloading decision of collaborative task execution between platoons and a MEC server. Both \cite{platoon_1, platoon_3} considered how to determine the location of task execution either on a vehicle, offloading to other platoon members, or an associated MEC server. However, \cite{platoon_1} focused on minimizing the offloading cost, while \cite{platoon_3} aimed at reducing the average energy consumption. 
The authors in \cite{platoon_2} proposed a federated offloading method that exploits horizontal offloading paths between vehicles, with the objective of minimizing total latency.
In \cite{2_tier_1}, the authors aimed at minimizing the power consumption of MEC servers and vehicles.
Zhao et al. recently utilized MEC and cloud computing resources simultaneously for offloading \cite{3-tier_1}. In that work, vehicles could offload their computation tasks to a MEC server or the cloud via RSUs. The objective was to maximize the system's utility by optimizing both the offloading strategy and resource allocation. In \cite{8402110}, Y. Lin et al. addressed the traffic and capacity allocation problem in a three-tier model and proposed an optimization algorithm consisting of two phases. The first was adjusting the capacity allocation, and optimizing the traffic allocation. Their objective was to minimize total capacity and guarantee that at least some traffic has satisfying latency constraints. Nguyen et al. in \cite{Khiem_Le_MEC} proposed a 3-tier offloading model that leverages both MEC and cloud computing. The authors first provided an explicit theoretical model to formulate the average task processing latency. They then proposed a meta-heuristic approach to determine sub-optimal offloading probabilities for the vehicles. 
In \cite{9679398}, the authors proposed an offloading protocol aiming at reducing communication costs while ensuring the packet latency constraint.

Unlike previous research, we use three communication planes simultaneously, namely, vehicle-to-cloud, vehicle-to-RSU, and vehicle-to-vehicle, with the goal of reducing 4G communication costs while maintaining information freshness.  

\section{Preliminaries} \label{sec:preliminaries}
In this section, we describe two techniques that will be used in our solution, namely Q-learning and Fuzzy logic. 

\subsection{Q-learning}
\begin{figure}[tb]
    \centering
    \includegraphics[width=0.55\columnwidth]{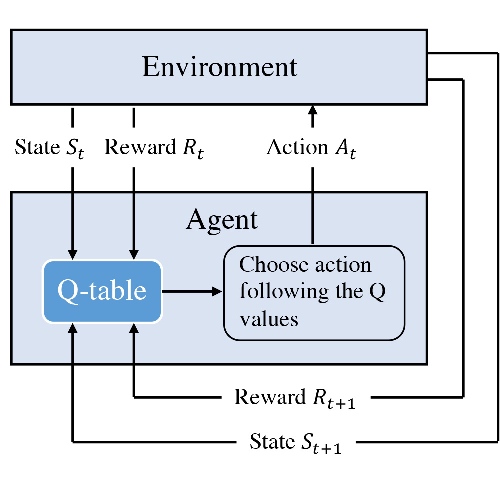}
    \caption{Q-learning overview.}
    \label{fig:q_learning}
\end{figure}

Q learning is a reinforcement learning technique that has been extensively utilized to tackle decision-making problems.
Reinforcement learning is mainly based on the trial-and-error paradigm.
A reinforcement learning framework, in particular, is comprised of five major components: environment, agent, action, state, and reward.
The agent performs an action at each state and interacts with the environment.
The environment then responds with a signal indicating the efficacy of the action. Finally, this goodness is quantified by a so-called award.
Based on the reward, the agent accumulates experiences from previous actions and progressively improves the actions to maximize the accumulative reward.
The Q learning model chooses actions based on the so-called Q values stored in a Q table. After each action, the agent updates the Q table using the following Bellman equation.
\begin{equation}
\small
\label{eq}
Q(S_t, A_t) \leftarrow (1 - \alpha)Q(S_t, A_t) + \alpha[\mathcal{R}_t + \gamma\max_{a}Q(S_{t+1}, a)],
\end{equation}
where, $S_t$ and $S_{t +1}$ denote the states at time slots $t$ and $t+1$, respectively; $a_t$ represents the action performed at time slot $t$, and $\mathcal{R}_t$, $Q(S_t, A_t)$ depict the reward and Q value when performing action $A_t$ at state $S_t$; $\max_{a}Q(S_{t+1}, a)$ is the maximum value that may be obtained for all possible actions $a$ at the next state $S_{t+1}$.
In addition, $\alpha$ and $\gamma$ are two hyperparameters named learning and discount rates, respectively.
These hyperparameters range from $0$ to $1$.
 
\subsection{Fuzzy logic}
\begin{figure}
    \centering
    \includegraphics[width=1.0\columnwidth]{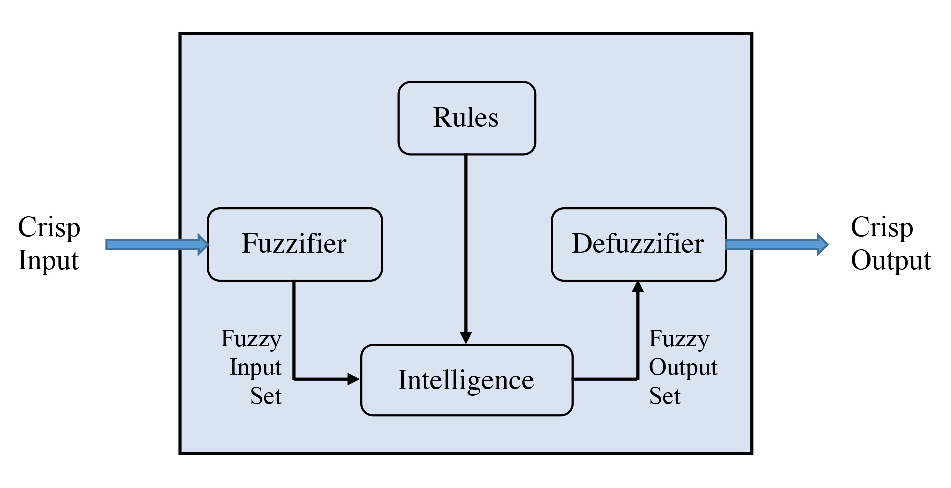}
    \caption{Fuzzy logic systems architecture.}
    \label{fig:fuzzy_logic}
\end{figure}

The Fuzzy logic \cite{zadeh1988fuzzy} architecture depicted in Fig. \ref{fig:fuzzy_logic} consists of four main components: Fuzzification module, Knowledge base, Inference engine, and Defuzzification module.

\subsubsection{Fuzzification Module} 
The fuzzification module converts the crisp values of the control inputs into fuzzy values. A fuzzy variable has values, which are defined by linguistic variables (fuzzy sets or subsets) such as low, medium, high, where each is defined by a gradually varying membership function.

\subsubsection{Knowledge Base} The knowledge base stores IF-THEN rules provided by experts. The expert knowledge is a collection of Fuzzy membership functions and a set of Fuzzy rules having the form: \textbf{IF} (conditions are fulfilled) \textbf{THEN} (consequences are inferred). More explicitly, a Fuzzy rule $R_i$ with $k$-inputs and $1$-output can be represented as follows.
\begin{equation}
\label{rule}
\begin{split}
    \mathfrak{R}_i: {} &\mathbf{IF} (I_1 \text{ is } A_{i1}) \Theta (I_2 \text{ is } A_{i2}) \Theta \ldots \Theta (I_k \text{ is } A_{ik})\\
    &\mathbf{THEN} (O \text{ is } B_i),
\end{split}
\end{equation}
where $\{I_1, \cdots ,I_k\}$ represents the crisp inputs to the rule. $\{A_{i1}, \cdots, A_{ik}\}$ and $B_i$ are linguistic variables. The operator $\Theta$ can be \textbf{AND}, \textbf{OR}, or \textbf{NOT}.

\subsubsection{Inference Engine} The inference engine deduces the Fuzzy control actions by employing Fuzzy implication and Fuzzy rules of inference. It calculates the membership degree ($\mu$) of the output for all linguistic variables by applying the rule set described in the \textit{Knowledge Base}. For Fuzzy rules with many inputs, the output calculation depends on the operators used inside it. The calculation for each type of operator is described as follows:

\begin{equation}
\begin{aligned}
    (I_i \ \mathrm{is} \ A_i \ & \mathbf{AND} \ I_j \ \mathrm{is} \  A_j): \\ 
    & \mu_{A_i \cap A_j} (I_{ij}) = \min (\mu_{A_i}(I_i), \mu_{A_j}(I_j)) \\
    (I_i \ \mathrm{is} \ A_i \ & \mathbf{OR} \ I_j \ \mathrm{is} \  A_j): \\
    & \mu_{A_i \cup  A_j} (I_{ij}) = \max (\mu_{A_i}(I_i), \mu_{A_j}(I_j)) \\
    (\mathbf{NOT} \ I_i & \ \mathrm{is} \ A_i): \\
    &\mu_{\bar{A_i}}(I_i) = 1 - \mu_{{A_i}}(I_i) \\
\end{aligned}
\end{equation}

\subsubsection{Defuzzification Module} The defuzzification module translates Fuzzy control values into crisp numbers, that is, it links a single point to a Fuzzy set, given that the point belongs to the support of the Fuzzy set. The most well-known defuzzification technique is the centre-of-area (COA) or centre-of-gravity (COG). 
For continuous membership function, the defuzzified value denoted as $x^*$ using COG is defined as:
\begin{equation}
\label{eqn:fuzzy_defuzzification}
    x^{*} = \frac{\int x.\mu_A(x)dx}{\int \mu_A(x)dx}, 
\end{equation}
where $\mu_A(x)$ is the output membership of the linguistic variable $A$.

\section{Fuzzy Q-learning-based Opportunistic Communication} \label{sec:fuzzy_q_learning}
\subsection{Network Model}
\begin{figure}
    \centering
    \includegraphics[width=1\columnwidth]{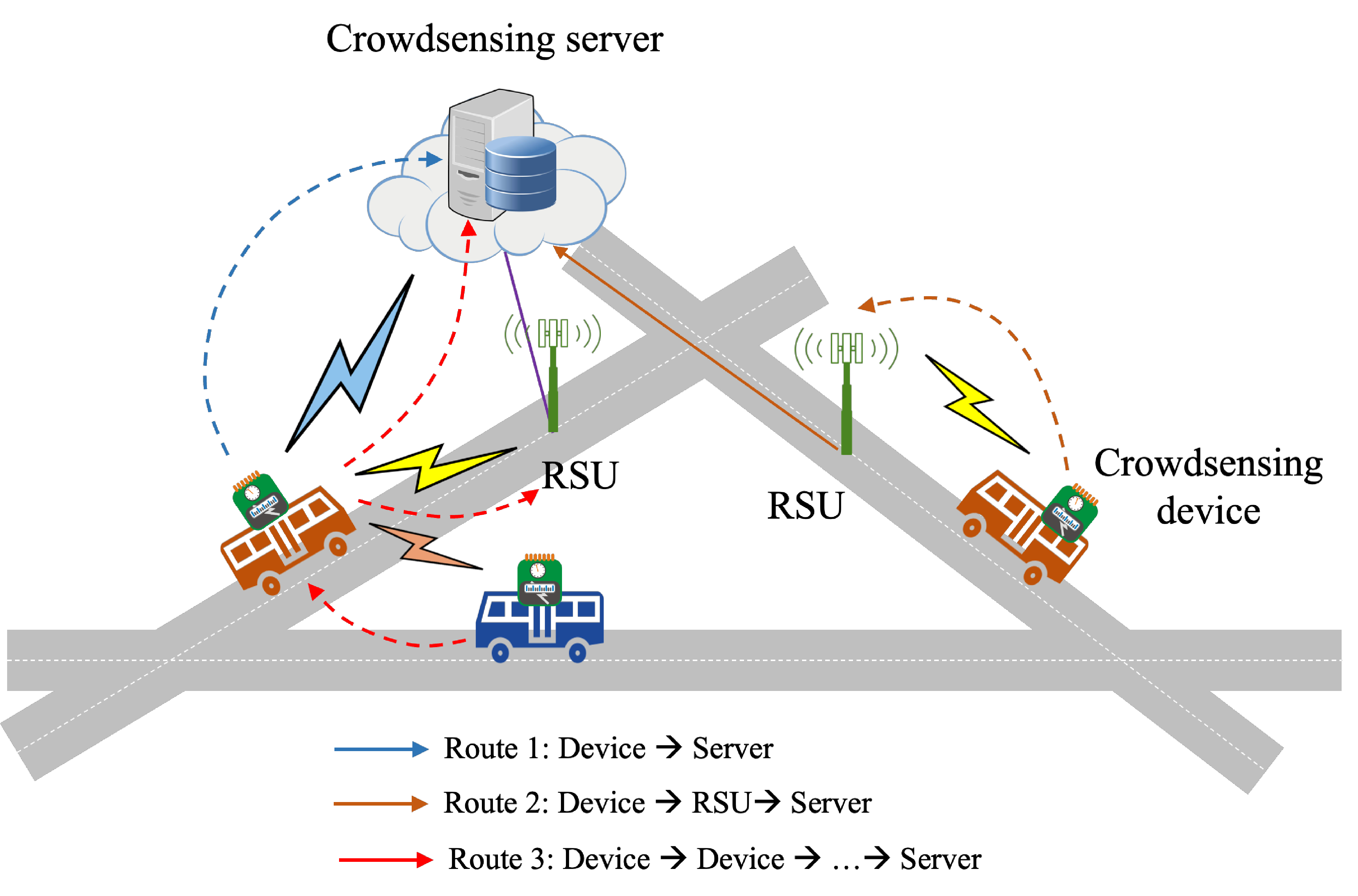}
    \caption{Network model.}
    \label{fig:network_model}
\end{figure}
Figure \ref{fig:network_model} depicts our network architecture, which comprises three parts: crowdsensing devices, a crowdsensing server, and roadside units.
Crowdsensing devices are sensory data-gathering devices that are installed on buses. They include 4G and Wi-Fi network interfaces.
The crowdsensing server, located in the cloud, is responsible for collecting data from devices.
RSUs are roadside processing units equipped with communication and computation capabilities.
The data acquired from crowdsensing devices will be sent to the crowdsensing server in one of three ways:
\begin{enumerate}
    \item The devices use a 4G communication channel to send sensory data directly to the server.
    \item The devices use Wi-Fi to relay data to the RSUs, the RSUs then deliver the data to the server through the wired network.
    \item Using a Wi-Fi channel, devices may communicate data to a neighboring device. The nearby device will then process the packet, i.e., it can be passed to the RSU, forwarded directly to the server, or forwarded to another device.
\end{enumerate}
We assume that the 4G channel is always accessible. Therefore the first transmission pathway is always feasible.
Furthermore, since Wi-Fi has limited communication range, the second and third actions can only be performed when the current device reaches the coverage area of another device or a RSU.
As shown in the Table \ref{tab:4Gcost}, the communication charge for 4G is much more than that of the wired network, and Wi-Fi is usually free.
Therefore, this study aims to propose an offloading protocol which leverages three transmission routes mentioned above, such that the total number of packets delivered by the 4G channel is the lowest while ensuring that data delay does not exceed a certain threshold $\delta$.
Here the term "data latency" refers to the time it takes for data to reach the server from when it is measured.
To ease the presentation, we use the following notations, hereafter.
We assume that there are $n$ crowdsensing devices that are mounted on $n$ buses. 
Each device $D_i (i = 1,\dots,n)$ has a computing capacity of $C^{*}_{i}$ and the transmission range of $r_{D_i}$. 
We also assume that there are $m$ RSUs denoted as $R_j (j = 1,\dots,m)$ which has the transmission range of $r_{R_j}$.

\begin{table}
    \centering
    \caption{Notions \label{tab:notions}}
    \begin{tabular}{|c|p{180pt}|}
    \hline
         \emph{Notion} & \emph{Description} \\
         \hline
         $n$ & the number of crowdsensing devices \\
         $m$ & the number of RSUs \\
         $D_i$ & the $i$-th crowdsensing device \\
         $r_{D_i}$ & the transmission range of $D_i$  \\
         $C^{*}_i$ & the maximum computing capacity of $D_i$\\
         $R_j$ & the $j$-th RSU \\ 
         $r_{R_j}$ & the transmission range of $R_j$ \\
         $\delta$ & the data latency threshold\\
         $\mathcal{S}_i$ & the state space of agent $D_i$\\
         $S_i(t)$ & the state at a time slot $t$ of agent $D_i$\\
         $\mathcal{A}_i$ & the action space of agent $D_i$\\
         $A_i(t)$ & the action taken by agent $D_i$ at a time slot $t$\\
         $\mu_i(t)$ & the timing agent $D_i$ generates the last data at a time slot $t$\\
         $c_i(t)$ & the remaining computing resource of $D_i$ at time slot $t$ \\
         $c_\mathcal{N}(t)$ & the remaining computing resource of the nearest device at time slot $t$ \\
         $\Delta_i(t)$ &  the time interval from when $D_i$ generates the last data until the time slot $t$,  $\Delta_i(t)= t - \mu_i(t)$\\
         $\Delta{c}(t)$ & the difference in remaining capacity of $D_i$ and nearest device at time slot $t$, $\Delta{c}(t) = c_{\mathcal{N}}(t) - c_i(t)$\\
         $\theta$ & the priority factor\\
         \hline
    \end{tabular}
\end{table}


\subsection{Q-learning based modeling} \label{subsec:Q_learning}
In this section, we first define the state and action space of the Q-learning-based model in Section \ref{subsec:state}, respectively. 
We then propose a novel reward function for the OCVC problem in Section \ref{Qlearning-rewardfunction}.
In our Q-learning-based model, the network is considered the environment, while each crowdsensing device is an agent.
We utilize the distributed approach where each monitoring device runs its own Q-learning-based model. To facilitate the reading, we summarize the notations in Table \ref{tab:notions}. 

\subsubsection{State and Action Space}
\label{subsec:state}
For each device $D_i$, the state at a time slot $t$ is a quadruple consisting of the following items: 
\begin{itemize}
  \item $\mu_i(t)$: the timing $D_i$ generates the last packet, i.e., the time from when the last packet is generated until the current time.
  \item $c_i(t)$: the computing resource of $D_i$ that is remaining at time slot $t$. 
  \item $c_\mathcal{N}(t)$: the remaining resource of the nearest device $\mathcal{N}$, if $\mathcal{N}$ is in the communication range of $D_i$.
  \item $\mathcal{N}_i^R(t)$: a binary variable indicating whether $D_i$ is in the communication range of a RSU.
\end{itemize}
Furthermore, the first three entries of a state are rounded as follows.
$\mu_i(t)$  is rounded in time step units.
$c_i(t)$ and $c_\mathcal{N}(t)$ are both rounded in Megabytes.
In this way, we have discretized the value of the state vector.
As a result, the state space is limited.
\label{subsec:action}
A crowdsensing device can conduct one of the following actions at each time slot $t$:
\begin{enumerate}
    \item[i)] Keeping the data in the local queue,
    \item[ii)] Sending the data directly to the crowdsensing server via 4G communication channel, 
    \item[iii)] Sending the data to the nearest RSU, if $D_i$ is in the communication range of an RSU,
    \item[iv)] Sending the data to the nearest device, if they are in the communication range of each other. 
\end{enumerate}
Because both the state and the actions obtain discrete values, the size of the Q table is fixed.
As a result, we may use the common sequential searching technique to retrieve an entry in the Q table.

\subsubsection{Reward Function} \label{Qlearning-rewardfunction}
We denote by $\mathcal{R}_i(t)$ the reward received when device $D_i$ performs action $A_i(t)$. 
Our goal is to minimize the total amount of data transmitted by 4G while guaranteeing that the data latency does not exceed a predefined threshold $\delta$.
For each type of action, the action's goodness is reflected by different indicator. Therefore, instead of define a an unique formula for the reward function, we break it down into multiple cases as follows.

\small
\begin{numcases}
{\mathcal{R}_i(t)=}
 -p & \text{, if} \ $c_i(t) = 0$ \nonumber\\
& \text{or} \ $\Delta_i(t) > \delta$ \label{eq5} \\
    \frac{\theta \times C^*_i - c_i(t)}{1 + \Delta_i(t)} & \text{, sending to the} \nonumber\\
    & \text{crowdsensing server} \label{eq1} \\ 
\frac{C^*_i - \theta \times c_i(t)}{1 + \Delta_i(t)} & \text{, sending to a RSU} \label{eq2} \\ 
\frac{\Delta{c}(t)}{[1 + \Delta_i(t)] \times \left | \Delta{c}(t) \right |} & \text{, sending to the} \nonumber\\
& \text{nearest device} \label{eq3_4} \\ 
0 & \text{, keeping in the}\nonumber\\
& \text{local memory} \label{eq6}   
\end{numcases}
\normalsize
where $\Delta_i(t) = t - \mu_i(t)$ is the time elapsed from when the data is collected, $\Delta{c}(t) = c_{\mathcal{N}}(t) - c_i(t)$ in that $c_{\mathcal{N}}(t)$ is the remaining resource of the nearest device, and $p$ is a significantly large positive number.
$\theta$ is a parameter in the range of $[0, 1]$ which we call \emph{priority factor}. The value of $\theta$ is determined by Fuzzy logic as will be described in Section \ref{fuzzy-logic-reward}.
The rationale behind the reward function is as follows. 
Firstly, when an action results in a resource exhaustion of the current device, or when the data's latency exceeds the threshold, the agent will be punished by a substantial negative reward according to Formula (\ref{eq5}).
Otherwise, the reward is calculated by Formulas from (\ref{eq1}) to (\ref{eq6}) depending on the action type.

The reward obtained when performing the action of sending a packet to the server is represented by Formula (\ref{eq1}).
As can be seen, this reward is inversely proportional to $c_i(t)$, the device's remaining resource.
It indicates that the lower $c_i(t)$, the bigger the reward, indicating that the action of sending to the server is encouraged.
In contrast, when $c_i(t)$ increases, the reward for sending a packet to the server decreases.
Moreover, when $c_i(t)$ is large enough (more than $\theta \times C^{*}_i$), the reward turns negative, and transmitting to the server is discouraged. It means that when the remaining resource is significant enough, the device will temporarily not need to transmit the packet to the server to save money on 4G.
The reward for transmitting a packet to an RSU is represented by Equation (\ref{eq2}). Similarly to formula (\ref{eq1}), the smaller the remaining resource, the more strongly urged is the action of transmitting to an RSU.
The reward for transmitting to an adjacent device is represented by Equation (\ref{eq3_4}). This reward is positive if $\Delta c(t)$ is greater than $0$, i.e., the neighbor's remaining resource is greater than the current device.
It implies that transmitting to an adjacent device is only recommended if the neighboring device has a greater remaining resource than the current device.
Finally, rewards for actions of sending to the server/RSU/other devices share a common term of $\frac{1}{1+\Delta_i(t)}$. It should be noted that this is inversely proportional to $\Delta_i(t)$. As a result, this term encourages the agent to decide on offloading tasks as quickly as feasible rather than holding the packet in local memory.

\begin{algorithm}[t]
\caption{Action selection and Q table update}
\label{algorithm:q-learning}
\textbf{Input}: $\alpha$: learning rate, $\gamma$: discount factor, $\epsilon$: a small number; current $Q$ table; $S_t$: current state. \\
\noindent \textbf{Output}: the next action; updated $Q$ table.
\begin{algorithmic}[1]
\State \textit{// Choose the next action};
\State $r \leftarrow$ uniform random number between $0$ and $1$;
\If {$r < \epsilon$}
    \State $A_t \leftarrow$ random action from the action space;
\Else 
    \State $A_t \leftarrow \underset{A_t}{\text{argmax}}~Q(S_t,A)$;
\EndIf
\State \textit{// Update Q table};
\State $S_{t+1} \leftarrow $ performing $A_t$;
\State $\theta \leftarrow $ calculated by Algorithm \ref{algorithm:fuzzy-logic};
\State $\mathcal{R}_t \leftarrow $ calculated by Formulas (5)-(9);
\State $Q(S_t, A_t) \leftarrow (1 - \alpha)Q(S_t, A_t)$
    
    ~~~~~~~~~ $ + ~\alpha[\mathcal{R}_t + \gamma\max_{a}Q(S_{t+1}, a)]$;

\end{algorithmic}
\end{algorithm}

\begin{algorithm}[t]
\caption{Fuzzy logic-based $\theta$ determination}
\label{algorithm:fuzzy-logic}
\textbf{Input:} $c_i(t)$: the remaining resource;\\
\hspace*{\algorithmicindent} ~~~~ $C_i^*$: the maximum computing capacity;\\
\hspace*{\algorithmicindent} ~~~~ $\Delta_i(t)$: the elapsed time;\\
\hspace*{\algorithmicindent} ~~~~ $\delta$: the data latency threshold.\\
\textbf{Output:} $\theta$.

\begin{algorithmic}[1]
\Function{FuzzyLogic}{$c_i(t), C_i^*, \Delta_i(t), \delta$}
\State $C \leftarrow  c_i(t) / C_i^*$;
\State $\Delta \leftarrow  \Delta_i(t) / \delta$;
\State \textit{// Fuzzification}
\State $\mu_\mathrm{L}(C) = \mathrm{Trapezoidal}(C, a_L, b_L, c_L, d_L)$;
\State $\mu_\mathrm{M}(C) = \mathrm{Trapezoidal}(C, a_M, b_M, c_M, d_M)$;
\State $\mu_\mathrm{H}(C) = \mathrm{Trapezoidal}(C, a_H, b_H, c_H, d_H)$;
\State $\mu_\mathrm{L}(\Delta) = \mathrm{Trapezoidal}(\Delta, a_L, b_L, c_L, d_L)$;
\State $\mu_\mathrm{M}(\Delta) = \mathrm{Trapezoidal}(\Delta, a_M, b_M, c_M, d_M)$;
\State $\mu_\mathrm{H}(\Delta) = \mathrm{Trapezoidal}(\Delta, a_H, b_H, c_H, d_H)$;
\State \textit{// Fuzzy controller}
\State $\mathrm{M} \leftarrow \textrm{null}$;
\For{$A \in \{L, M, H\}$}
    \For{$B \in \{L, M, H\}$}
        \State $\mu = \mathrm{min}\{\mu_A(C), \mu_B(\Delta)\}$;
        \State $\mathrm{M}.\textrm{add}(\mu)$;
    \EndFor
\EndFor
\State \textit{// Defuzzification}
\State $l = \underset{\mathrm{\mu}}{argmax}\mathrm {M}$;
\State $\mathrm{D} \leftarrow \textrm{the output of rule } \mathfrak{R}_l$;
\State $\theta \leftarrow \textrm{the value of } \theta \textrm{ by CoG function}$;
\State
\Return $\theta$
\EndFunction
\end{algorithmic}
\end{algorithm}

To choose the next action, we leverage the $\epsilon$-greedy policy.
In particular, the $\epsilon$ steadily declines over time as follows:
\begin{equation}
        \epsilon \leftarrow \epsilon \times \max \left \{0, \frac{\text{maximum time} - \text{current time}}{\text{maximum time}}\right \},
\end{equation} 
where the maximum time is a predefined threshold.
The details of this policy are described in Algorithm \ref{algorithm:q-learning}.
At each step, the agent selects a random action with a probability of $\epsilon$ from the action space (line 4 in Algorithm \ref{algorithm:q-learning}) and the action with the greatest Q value with a probability of $1-\epsilon$ (line 6 in Algorithm \ref{algorithm:q-learning}).

\subsection{Fuzzy Logic-based priority factor determination} \label{fuzzy-logic-reward}

\subsubsection{Motivation\label{fuzzy_motivation}}
\begin{figure*}
\minipage{0.64\textwidth}
  \minipage{0.5\textwidth}
  \includegraphics[width=\linewidth]{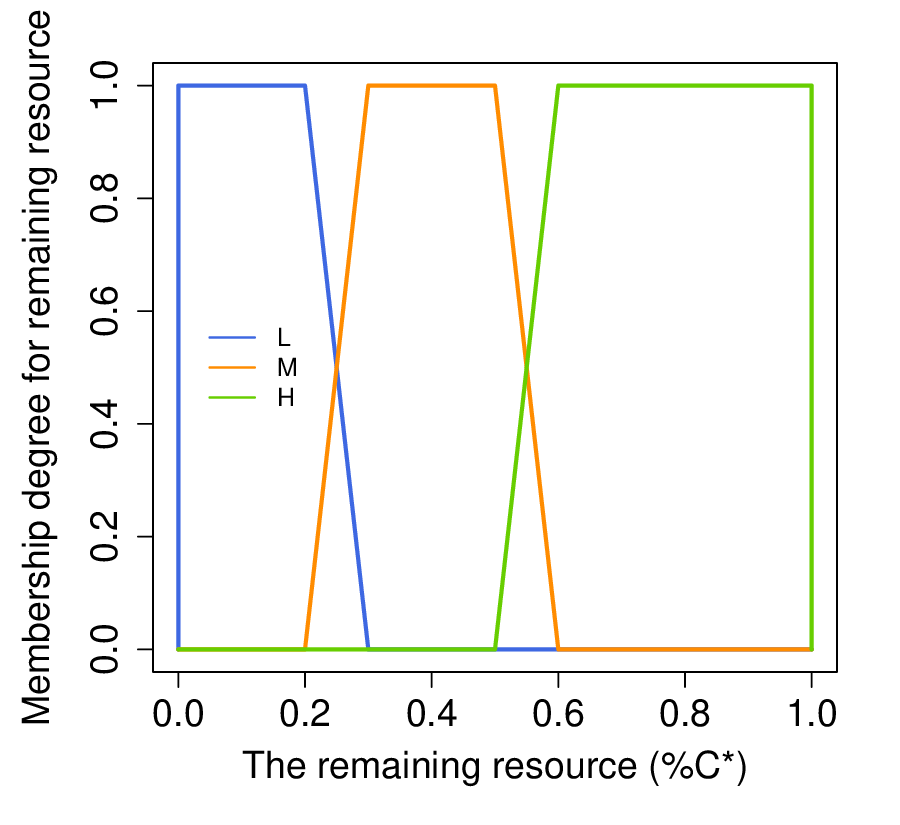}
  \endminipage
  \minipage{0.5\textwidth}
  \includegraphics[width=\linewidth]{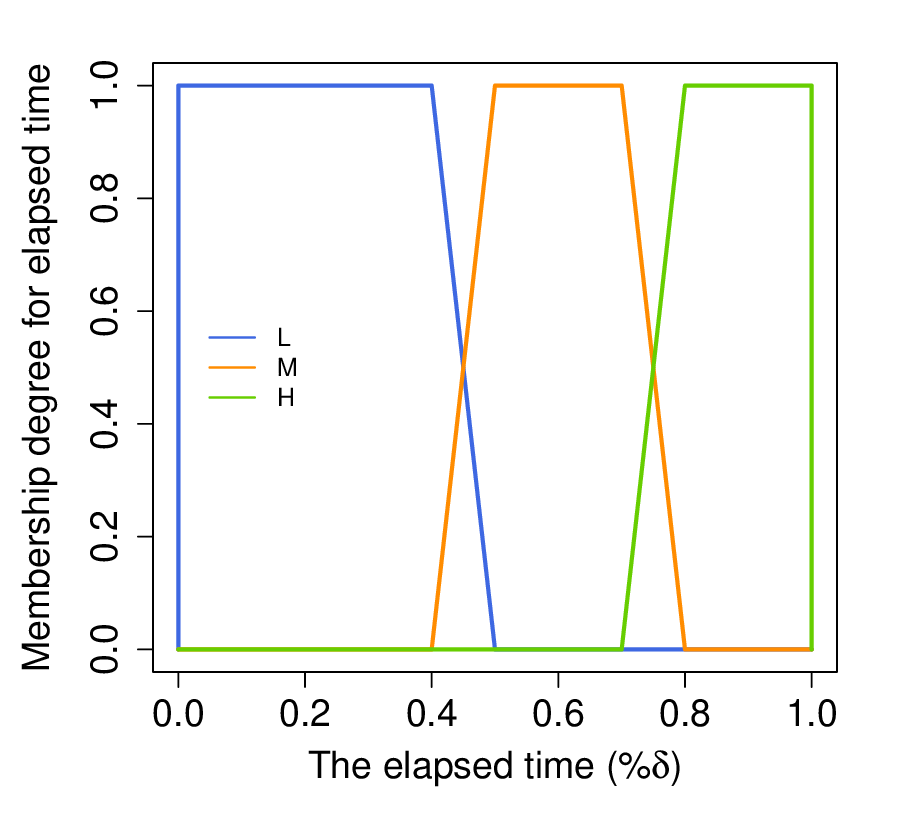}
  \endminipage
  \caption{Fuzzy input membership function with three linguistic variables: \emph{low} (L), \emph{medium} (M) and \emph{high} (H).}\label{fig:fuzzy-input}
\endminipage\hfill
\minipage{0.32\textwidth}%
  \includegraphics[width=\linewidth]{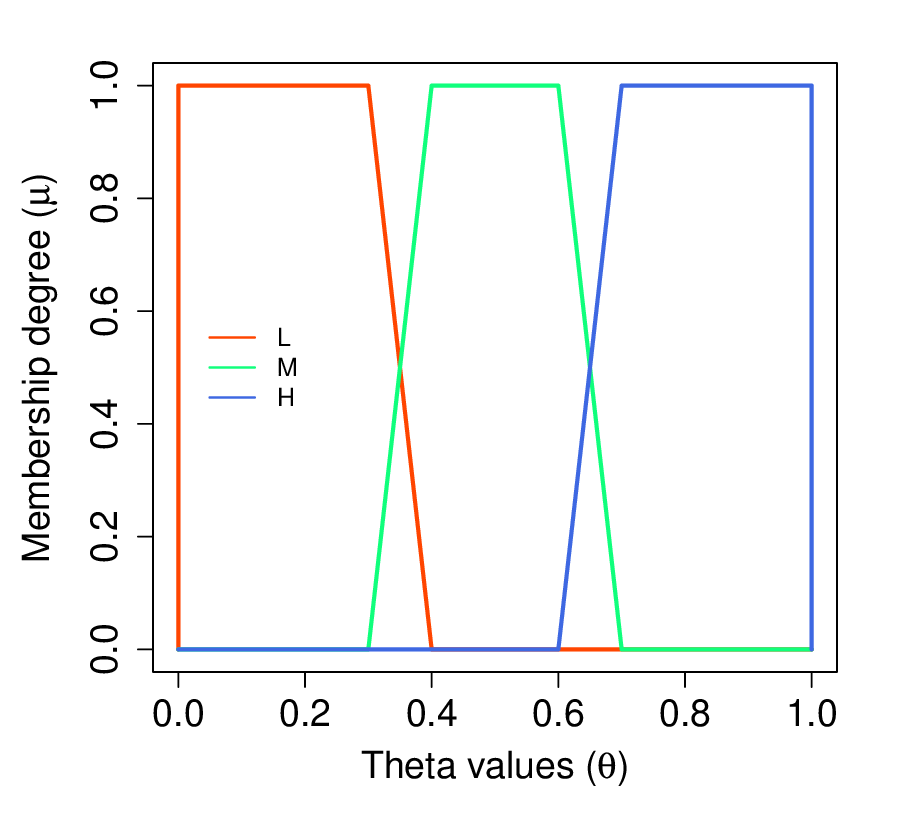}
  \caption{Fuzzy output membership function.}\label{fig:fuzzy-output}
\endminipage
\end{figure*}

We found that simultaneously optimizing two objectives, namely minimizing the total amount of data transmitted over the 4G communication channel and guaranteeing the data latency, is challenging. 
Specifically, devices tend to store data until they can send it to an RSU to reduce the usage of 4G transmission. 
This strategy, however, may cause a large delay due to the waiting period.
As a result, rather than using a fixed value for $\theta$, we propose a mechanism for adjusting $\theta$ dynamically in response to network status.
We observe that the remaining resource and the amount of time the data has elapsed are two factors that influence a device's behavior.
The device may tolerate more latency when the device's remaining resource is large or data latency is minimal; therefore, the agent should avoid transmitting immediately to the server to save on communication expenses.
As a result, $\theta$ should be insignificant when the device's remaining resource is large and the data latency is low.
In contrast, when the remaining resource in the device is low or data latency is large, the device should send data to the server rather than transferring it to the nearest device to ensure the latency constraint is satisfied.
Consequently, $\theta$ should be set high enough that the reward of the action sending to the server surpasses the reward of relaying to the nearest device.
Furthermore, as long as the device is within an RSU's communication range, transmission to the RSU takes precedence, i.e., the action of sending to an RSU is designed to get the highest reward of all the actions.

Motivated by the observation mentioned above, we design a Fuzzy-based algorithm for dynamically adjust the value of $\theta$. The pseudo-code of the algorithm is presented in Algorithm \ref{algorithm:fuzzy-logic}.

\subsubsection{Fuzzification}

The input of the fuzzification process is a pair consisting of the remaining resource, and the latency the data has elapsed (i.e., defined by the time from when the data is measured until the current time). 
The input then is mapped to three linguistic variables: \emph{low} (L), \emph{medium} (M) and \emph{high} (H).
The output of the Fuzzification is also mapped to three respective levels: \emph{low}, \emph{medium} and \emph{high}. 
We use the trapezoidal Fuzzy set that could be described by the following formula:
\begin{equation}
\textrm{Trapezoidal}(x, a, b, c, d) = \begin{cases}
0 &  \text{ if } x \leq a \\ 
\frac{x - a}{b - a} & \text{ if }  a \leq  x \leq b \\ 
1 & \text{ if }  b \leq x \leq  c \\ 
\frac{d - x}{d - c} &  \text{ if } c \leq x \leq d \\
0 & \text{ if } d \leq x,
\end{cases}
\end{equation}
where $x$ is the input, and $a, b, c, d$ are parameters. 
The values of $a, b, c, d$ are represented in Table \ref{tab:fuzzy_input_variables} and  \ref{tab:fuzzy_output_variable}. 
Fig.\ref{fig:fuzzy-input} and  \ref{fig:fuzzy-output} illustrate the shape of the input membership functions.

\begin{table}
    \centering
    \caption{Input variables with their linguistic values and corresponding membership function}
    \label{tab:fuzzy_input_variables}
    \begin{tabular}{|p{60pt}|c|c|}
        \hline
        Input variable & Linguistic value & Membership function \\
        \hline
        \multirow{3}{*}{\shortstack{The remaining \\ resource ($\%C^*$)}} & low (L) &  [0, 0, 0.2, 0.3] \\
        \cline{2-3}
        & medium (M) & [0.2, 0.3, 0.5, 0.6] \\
        \cline{2-3}
        & high (H) & [0.5, 0.6, 1, 1] \\
        \hline
        \multirow{3}{*}{\shortstack{~The elapsed \\time ($\%\delta$)}} & low (L) & [0, 0, 0.4, 0.5] \\
        \cline{2-3}
        & medium (M) & [0.4, 0.5, 0.7, 0.8] \\ 
        \cline{2-3}
        & high (H) & [0.7, 0.8, 1, 1] \\
        \hline 
    \end{tabular}
\end{table}
\begin{table}
    \centering
    \caption{Output variable with their linguistic values and corresponding membership function}
    \label{tab:fuzzy_output_variable}
    \begin{tabular}{|c|c|c|}
        \hline
        Output variable & Linguistic value & Membership function \\
        \hline
        \multirow{3}{*}{ $\theta$ } & low (L) &  [0, 0, 0.3, 0.4] \\
        \cline{2-3}
        & medium (M) & [0.3, 0.4, 0.6, 0.7] \\ 
        \cline{2-3}
        & high (H) & [0.6, 0.7, 1, 1] \\
        \hline
    \end{tabular}
\end{table}

\subsubsection{Knowledge Base}
We have two input variables, namely the remaining resource, and the elapsed time, each of which is transformed into three Fuzzy sets. 
Therefore, we have $3^2 = 9$ Fuzzy rules in total. 
The rules are shown in Table \ref{tab:fuzzy_rules}. 
These rules are designed based on the observation described in Section \ref{fuzzy_motivation}.
Our rules have the form of ``\textbf{IF} (\textit{the remaining resource} is A) \textbf{AND} (\textit{the elapsed time} is B) \textbf{THEN} ($\theta$ is C)", where $A, B, C$ obtain the values of \emph{low}, \emph{medium}, and \emph{high}.
To ease the presentation, we denote \emph{low}, \emph{medium}, and \emph{high} as $L$, $M$, and $H$, respectively. 

\begin{table}
    \centering
    \caption{Fuzzy rules}
    \label{tab:fuzzy_rules}
    \begin{tabular}{|c|c|c|c|}
        \hline
        \multirow{2}{*}{No.} & \multicolumn{2}{c|}{Input} & Output \\
        \cline{2-4}
        & The remaining resource & The elapsed time & $\theta$ \\
        \hline
        1 & L & L & M\\
        \hline
        2 & L & M & H\\
        \hline
        3 & L & H & H\\ 
        \hline
        4 & M & L & L\\
        \hline
        5 & M & M & M\\ 
        \hline
        6 & M & H & H\\ 
        \hline
        7 & H & L & L\\
        \hline
        8 & H & M & L\\
        \hline
        9 & H & H & M\\ 
        \hline
    \end{tabular}
\end{table}

\subsubsection{Inference Engine}
As our Fuzzy rules are based on \textbf{AND} operator, the output membership function is defined as below.
\begin{equation}
\label{eqn:fuzzy_inference} 
\begin{aligned}
    \mu_i = \min\{ & \mu_A(\text{the remaining resource}),\\ &\mu_B(\text{the elapsed time})\}, \forall i = 1,\cdots,9.
\end{aligned}
\end{equation}

\subsubsection{Defuzzification}
After going through the steps above, the Fuzzy set with the highest membership degree is considered as the output variable. Finally, we utilize the \textit{CoG} function in Formula (\ref{eqn:fuzzy_defuzzification}) to calculate the crisp value of the output's fuzzy set.

\subsection{Computational Complexity} 

We analyze the computational complexity to choose the next action and update the Q table in the following.
Since the agent utilizes the Epsilon Greedy policy to choose the next action, it needs to check the Q values of all actions in the action space.
The computational complexity of this operation is $O(|\mathcal{A}|)$, where $\mathcal{A}$ indicates the action space.
To update the Q table, the agent first utilizes Fuzzy Logic to determine the value of $\theta$ (Algorithm \ref{algorithm:fuzzy-logic}), which has a computational complexity of $O(\mathrm{L}_{C} \times \mathrm{L}_{\Delta})$, where $\mathrm{L}_{C}$ and $\mathrm{L}_{\Delta}$ are the number of linguistic values of the remaining resource and the elapsed time, respectively.
The agent then calculates the reward and uses Formula (\ref{eq}) to update the Q value. This operation has a computational complexity of $O(|\mathcal{A}| \times |\mathcal{S}|)$, where $\mathcal{S}$ denotes the state space.
As a result, the total time for updating the Q table at each step is $O(|\mathcal{A}| + |\mathcal{A}| \times |\mathcal{S}| + \mathrm{L}_{C} \times \mathrm{L}_{\Delta})$, which is equivalent to $O(|\mathcal{A}| \times |\mathcal{S}| + \mathrm{L}_{C} \times \mathrm{L}_{\Delta})$.

\section{Evaluation} \label{sec:experimental}
\subsection{Methodology}
\label{sec:methodology}

In this section, we evaluate the efficiency of our proposed algorithm in terms of optimizing the communication performance and cost.

\textbf{Simulation model:} 
We developed an in-house simulator in Python programming language version 3.8.8. The packet transmission process is implemented by extending the queue-model proposed in~\cite{Khiem_Le_MEC}. Our code can be access via \cite{simulator}. 
In which, the devices iteratively generate homogeneous packets of the same size. 
We denote by $\lambda_d$ the packet generation interval
After being created, packets are queued in the device and waiting for transmission. If the queue is full, the packet will be dropped.
As a result, data latency comprises two parts: waiting time in the queue and transmission time from the device to the server.
The transmission time is proportional to the packet size and inversely proportional to the communication channel bandwidth.
At each time slot, the device picks a packet from the queue and transmits it in one of three ways:
\begin{itemize}
    \item Using 4G to communicate directly to the server.
    \item Transferring data to an RSU through Wi-Fi and then transmitting data from the RSU to the server via the wired network.
    \item Relaying to a nearby device via Wi-Fi and then following that device's policy (transmitting directly to the server, transferring to RSU, or continuing to forward to another device) to send the packet to the server.
\end{itemize}
The offloading method is determined by the algorithms (our algorithm and the baselines). If the device chooses the second or third offloading mode but is not within the coverage of the RSU (or another device), it will hold the packet in the queue and wait for the next time slot (if the queue is not full), or it will transmit it immediately to the server using 4G (otherwise). We refer to the first case as \emph{offload-hit} and the second case as \emph{offload-missed}.

\textbf{Simulation environment:} In this work, we use the data collected within two days of bus routes in Seattle City, Washington \cite{jetcheva2003design} to simulate the movement of vehicles. Each data point includes the time and position of a bus. 
We only collect data of buses whose active time is no less than 90 minutes per day.
We then generate the RSUs' position on the map along each bus route. In which, the RSUs are concentrated in the city center, 1 - 3 km apart, while in the suburbs there will be a sparser number of RSUs, 4 - 8 km apart. 

\textbf{Evaluation metrics:} It is worthy to note that the target of this work is to keep the number of packets that can reach the server as much as possible (i.e., \textit{maximizing the delivery ratio}), while reducing the amount of data with a long latency (i.e., \textit{guaranteeing the information freshness}) and lessening the 4G communication usage rates (i.e., \textit{minimizing the communication cost}). 
Although we have already taken the wired and Wi-Fi costs into account since the per-second communication costs of these two plans are negligible compared to the cost of 4G communication, we focus only on the 4G communication ratio in this section.
First, we define the rate of dropped packets, called $r_{drop}$, for the delivery ratio.
This metric is computed by dividing the number of dropped packets by the number of packets transmitted.
To analyze the second criteria, ensuring the packet's freshness, we propose the term $\delta$-delayed packets, calculated by the total number of packets with latency higher than a threshold $\delta$.
In addition, we also define an additional term \textit{rate of $\delta$-delayed packets} (denoted $r_{delay}$), which is the ratio of the $\delta$-delayed packets to the total number of packets sent.
Finally, to evaluate the communication cost, we measure the proportion of packets sent using 4G out of the total number of packets. This number is named $r_{server}$\footnote{Assumption: the cost of sending a packet via the 4G communication channel is much higher than that of using Wi-Fi or wired network as an example shown in Table~\mbox{\ref{tab:4Gcost}}}.
In addition, we also define $r_{rsu}$, which is the rate of packets transferred via relay from RSU to the server.
 
\begin{table}
    \centering
    \caption{Simulation parameters \label{tab:params}}
    \begin{tabular}{| l | c |}
    \hline
         \emph{Parameter} & \emph{Value} \\
         \hline
         Packet size & 1 Mb\\
         RSU transmission range & 250 Meter \\ 
         Sensor transmission range & 40 Meter \\
         RSU-server's link bandwidth (wired network) & 10 Gbps \\
         Sensor-RSU's link bandwidth (wifi network) & 1 Gbps \\
         Sensor-server's link bandwidth (4G communication)  & 500 Mbps \\
         The length of a time slot ($\mathcal{T}$) & 1 Min \\ 
         Packet generation interval at a sensor ($\lambda_d$) & 1 $\sim$ 5 $\mathcal{T}$ \\
         Data latency threshold ($\delta$) & 5 $\sim$ 25 $\mathcal{T}$ \\
         Sensor's computing capacity $C^*$ & 25 Mb \\ 
         Number of RSUs ($m$) & 384 \\ 
         Number of vehicles ($n$) & 776 \\ 
         \hline
    \end{tabular}
\end{table}
\begin{figure*}[t]
    \centering
    \subfigure[$P_{keep}=0.1$]{\includegraphics[width=0.2\linewidth]{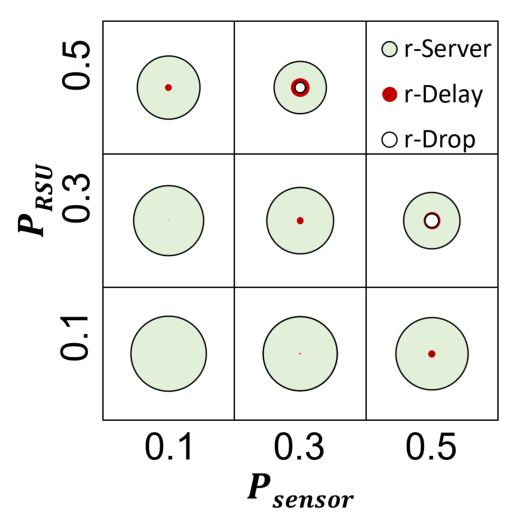}}
    \subfigure[$P_{keep}=0.3$]{\includegraphics[width=0.2\linewidth]{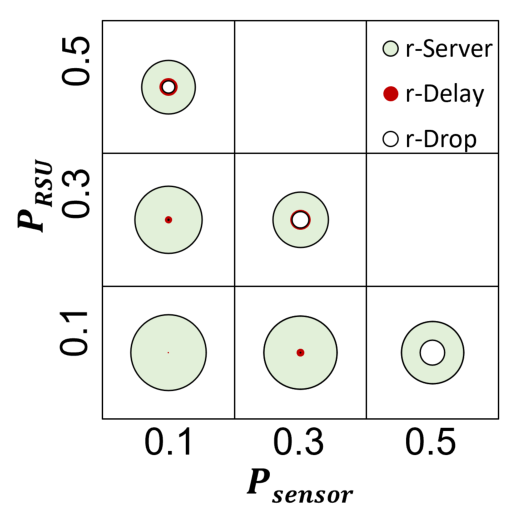}}
    \subfigure[$P_{keep}=0.5$]{\includegraphics[width=0.2\linewidth]{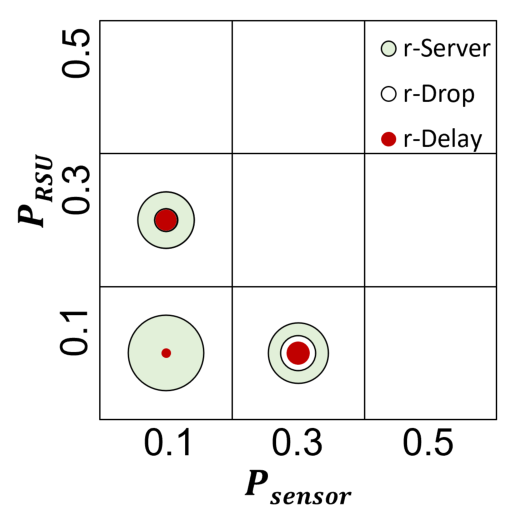}}
    \caption{Finding sub-optimal combination of $P_{keep}$, $P_{server}$, $P_{rsu}$ and $P_{sensor}$ for the FP baseline using grid search. $\delta$ is fixed to $10$ time slots. The bigger circle refers to the higher rate.}\label{fig:gridsearch} 
\end{figure*}

Reducing the energy consumption of crowdsensing devices (sensors) is well-known in the literature as one of the most critical challenges.
To understand how energy efficient these strategies are, we measure the total energy consumption of sensors employing different offloading strategies in this study.
We solely assess the energy consumption of communication because the variation in how the sensors perform with different offloading strategies in our simulation is in the communication method. Specifically, there are two types of communications at a given sensor, i.e., communication via WiFi and 4G channels. At a given sensor $i$, for a given packet $j$ transmitted via a given channel, we use the average power consumption ($PW_{j}$) of that channel (stated in Table~\ref{tab:energyCost}) and the simulated transmission latency of this packet ($T_j$) on the corresponding channel to estimate the energy consumption. That is $E = \sum_{i=1}^{n}{\sum_{j=1}^{N_i}{PW_{j} \times T_j}}$ where $n$ is the number of sensors and $N_i$ is the number of packet processed by this sensor.

\textbf{Comparison baseline:} Because there is no current work that handles the same problem as ours, to show the efficiency of our proposed method, we compare it with two baselines:

The first comparison baseline is a simple yet effective greedy opportunistic communication method defined below.
When a new packet is generated, the device performs an action based on the following rule:
\begin{itemize}
    \item If the device is in the communication range of an RSU, it always sends the packet to the RSU. The RSU will send the packet to the cloud server after that. 
    \item Otherwise, if the device is in the communication range of other devices nearer to an RSU than itself, it relays the packet to the new device.
    \item If the device does stays inside the communication range of neither an RSU nor another device, it sends the packet directly to the cloud server. 
\end{itemize}

The second baseline is a naive offloading strategy named \textbf{FP}. In FP, at a given time slot, a packet is randomly decided between four actions, namely keeping at the local, sending directly to the server, transferring to an RSU, and relaying to the nearest device, with fixed possibilities of $P_{keep}$, $P_{server}$, $P_{rsu}$ and $P_{sensor}$, respectively. 
We first conduct a grid search over the four parameters $P_{keep}$, $P_{server}$, $P_{rsu}$ and $P_{sensor}$ to find the sub-optimal combination. We then compare the performance of our proposed solution to the \textbf{FP}'s sub-optimal configuration.

In the following, we first compare the performance and cost of our proposed method with the baseline concerning particular settings of the packet generation interval $\lambda_d$ and the latency threshold $\delta$ in Section \ref{sec:compare}.
We then investigate the impacts of $\lambda_d$ and $\delta$ in Section \ref{sec:discussion}. 

\subsection{Comparison of the proposed method and the baseline}
\label{sec:compare}
In this comparison, we set the data latency threshold $\delta$ to $5$ and $10$ and the packet generation interval $\lambda_d$ to $1$. 
The remaining simulation parameters are derived from \cite{Khiem_Le_MEC}, and presented in  Table \ref{tab:params}.

\subsubsection{Sub-optimal configuration of FP}
In this section, we perform grid search on $P_{keep}$, $P_{rsu}$ and $P_{sensor}$ to identify the sub-optimal combination\footnote{$P_{server} = 1 - P_{keep} - P_{rsu} - P{sensor}$}. Fig.~\ref{fig:gridsearch} illustrates the rate of dropped packets ($r_{drop}$), $\delta$-delayed packets ($r_{delay}$) and the 4G communication ratio $r_{server}$ when we varies the value from 0.1 to 0.5.  The bigger circle refers to the higher rate. Through this result, we consider different configurations of FP that could have small packet dropped rate, e.g., $r_{drop} \approx 0$. We then pick up three configurations with the highest, medium and lowest 4G communication ratio $r_{server}$ to be the representations for the FP baselines (named as FP1, FP2, and FP3, respectively). The detail settings of FP are summarized in Table~\mbox{\ref{tab:prob}}. 

\begin{table}[t]
     \centering
     \caption{Configuration of the fix-possibility strategies}
     \label{tab:prob}
     \begin{tabular}{|c|c|c|c|c|c|}
         \hline
         Strategies & $P_{keep}$ & $P_{server}$ & $P_{rsu}$ & $P_{sensor}$\\
         \hline
         FP1 &0.1 & 0.7 & 0.1 & 0.1\\
         \hline
         FP2 &0.1 & 0.5 & 0.3 & 0.1\\ 
         \hline
         FP3 &0.1 & 0.3 & 0.5 & 0.1\\
         \hline
     \end{tabular}
\end{table}
\subsubsection{Delivery Ratio}
We shows the breakdown of the simulation packets (in percentage) of the proposed method as well as the greedy, and the FP baseline strategies in Fig.~\ref{fig:breakdown}. We consider the lower value to be the better performance.
First of all, with the FP strategy, devices tend to hold the data in their queue until they can send it to an RSU/or the nearest device (the \textit{offload-hit} as mentioned in section~\ref{sec:methodology}). This strategy leads to a longer delay of a message and a higher number of dropped packets (when the queue is full).
As a result, the FP baseline strategies can not avoid the packet-dropped issue in all the cases, e.g., 0.15\% of $r_{drop}$ as in FP3 when we set the latency threshold to 5 time slots.
In contrast, the greedy strategy does not keep the packet in the queue of a given sensor when there exist no RSUs and devices around it, yet encourage to send data directly to the cloud server. Thus, it could deliver all the packets to the server with a zero-packet dropped rate. Similar to the greedy strategy, our proposed method also achieves the zero-packet dropped rate and guarantees all the generated packets can reach the server. It is the result of flexibly constructing the priority factor $\theta$ using the Fuzzy logic approach which considers both the number of packets in the local queue and the data latency of packets.

\subsubsection{Information Freshness and Communication Cost}
The result in Fig.~\ref{fig:breakdown} also shows that our method provides a smaller number of $\delta$-delayed packet than the FP3 strategy, e.g., 1.86\% and 0.93\% which are $1.3\times$ and $2.94\times$ lower than those of FP3. 
In addition, although the $r_{delay}$ of the greedy, FP1, and FP2 strategies are lower than those of our proposed method, they require much more 4G cost than ours. Specifically, in all the experiments, our method requires the lowest communication cost ($r_{server}$), i.e., only around 60\% of packets travel through the 4G network while those are around $70\%$ as in the greedy. 
Furthermore, the 4G communication cost, i.e., $r_{server}$ of the FP strategies are much higher than expected, i.e., around $P_{server}$, due to the \textit{offload-missed} issue mentioned in Section~\ref{sec:methodology}.
When a device decides to send a packet to the nearest RSU/device but there exist no RSUs/devices around, or when the local queue is full, it sends the packet to server directly vía the 4G network. Such offload-missed issue increases the 4G communication cost significantly.
For example, in the FP1, FP2 and FP3, more than $96\%$, $86\%$ and $73\%$ of packets use the 4G communication channel (although it is designed with a fixed possibility of 70\%, 50\% and 30\%, respectively). 

\subsubsection{Energy Consumption}
\begin{figure}[t]
    \centering
    \includegraphics[width=0.47\textwidth]{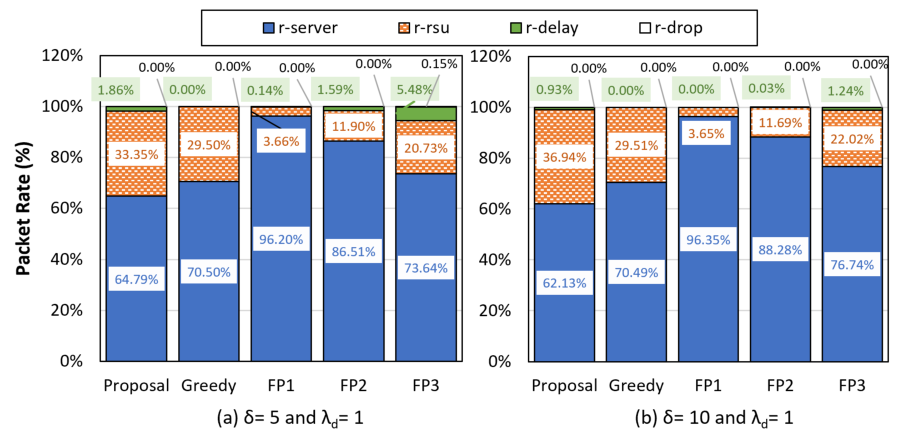}
    \caption{Relative breakdown of the simulation packets.}\label{fig:breakdown} 
\end{figure}

\begin{figure}
    \centering
    \includegraphics[width=0.48\textwidth]{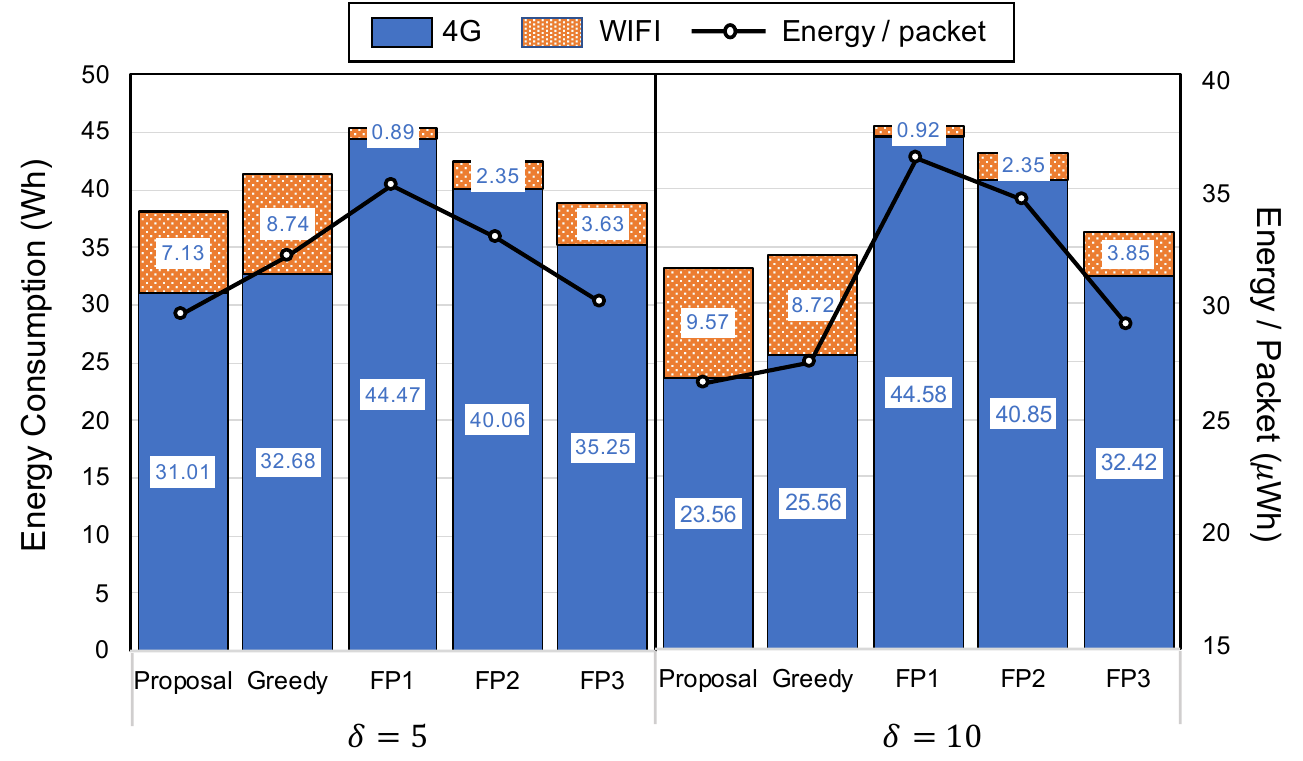} 
    \caption{Energy consumption of estimated offloading strategies ($\lambda_d$ =1).}
    \label{fig:Energy_Consumption}
\end{figure}

In Fig.~\ref{fig:Energy_Consumption}, we present the total energy consumption of offloading strategies using the stacked bars. The line series shows the average energy consumption of packets that reach the crowdsensing server.
Interestingly, in the case of  $\delta=5$, although the energy consumption of our proposed strategy is higher than that of FP3, the average energy consumption per packet of our proposal is smaller than that of FP3. 
It refers to the result in Fig.~\ref{fig:breakdown}(a) when
the rate of packets transmitted to the server (1 - $r_{drop}$) of our proposed strategy is higher while the rate of the packet transmitted to the server via the 4G channel ($r_{server}$) is smaller than those of the FP1 and FP2.
On the other hand, when the packets transmitted to the server of the four targeted strategies are similar, e.g., $\delta=10$, our proposal consumes the lowest energy because it uses the 4G channel least. 
\subsubsection{Summary}
\begin{figure*}
    \centering
   \includegraphics[width=0.5\textwidth]{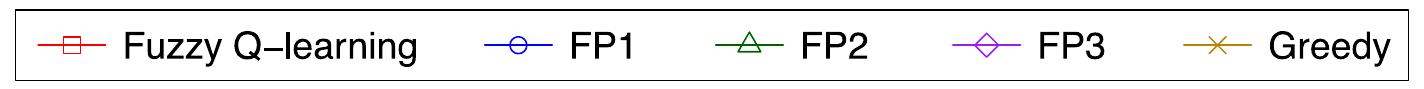}\\
    \subfigure[4G communication ratio]{\includegraphics[width=0.3\textwidth]{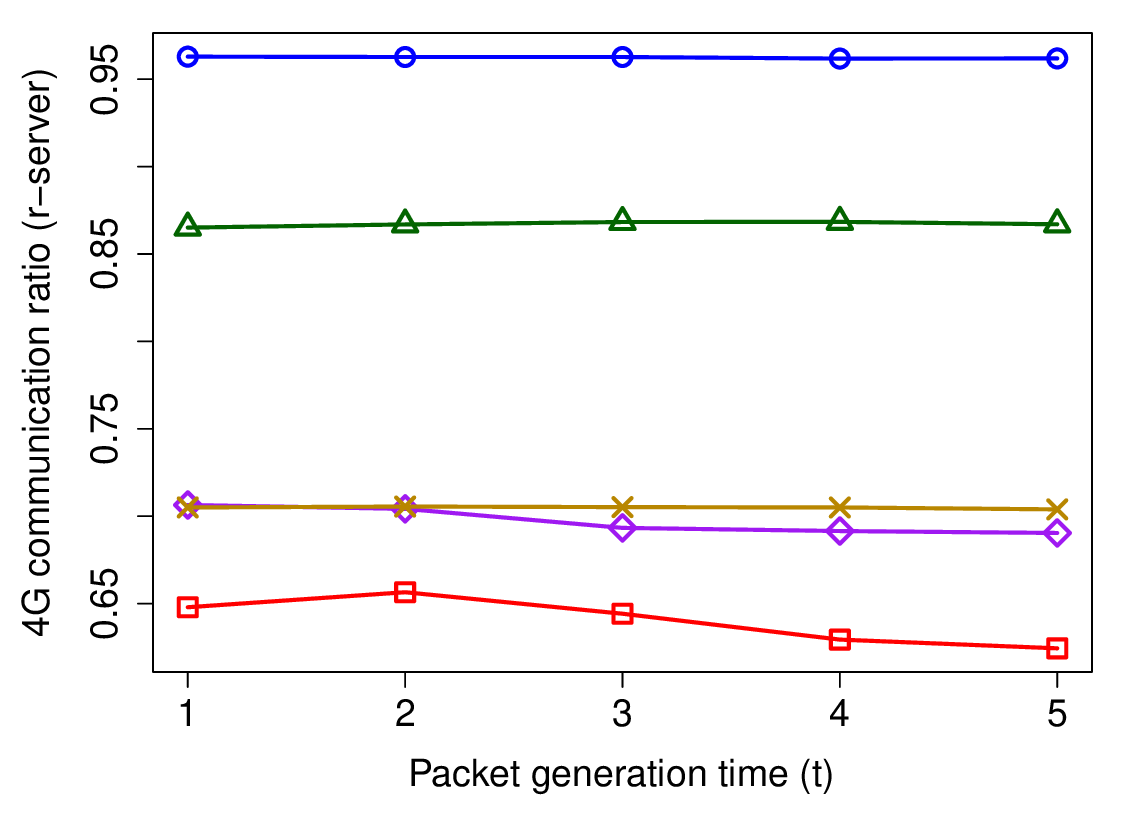}\label{fig:generation_time_1}}
    \hfill
    \subfigure[Rate of $\delta$-delayed packets]{\includegraphics[width=0.3\textwidth]{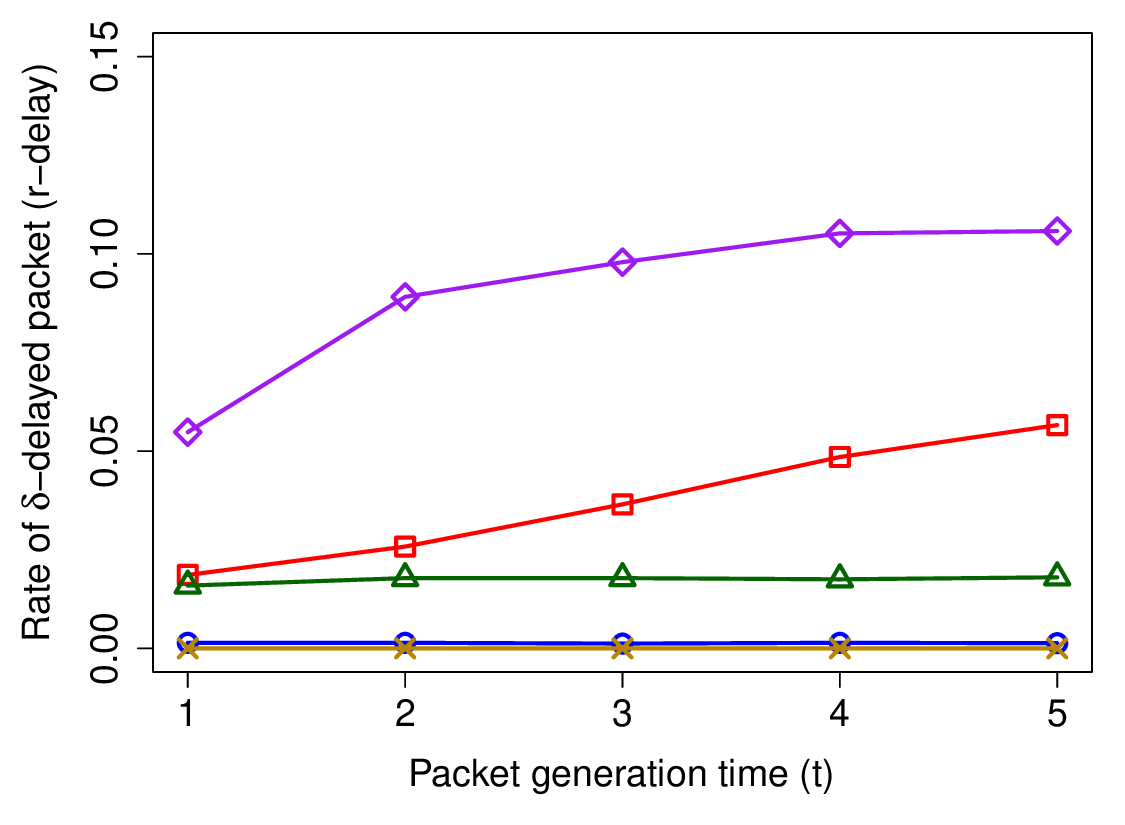}\label{fig:generation_time_2}}
     \hfill
    \subfigure[RSU to server rate]{\includegraphics[width=0.3\textwidth]{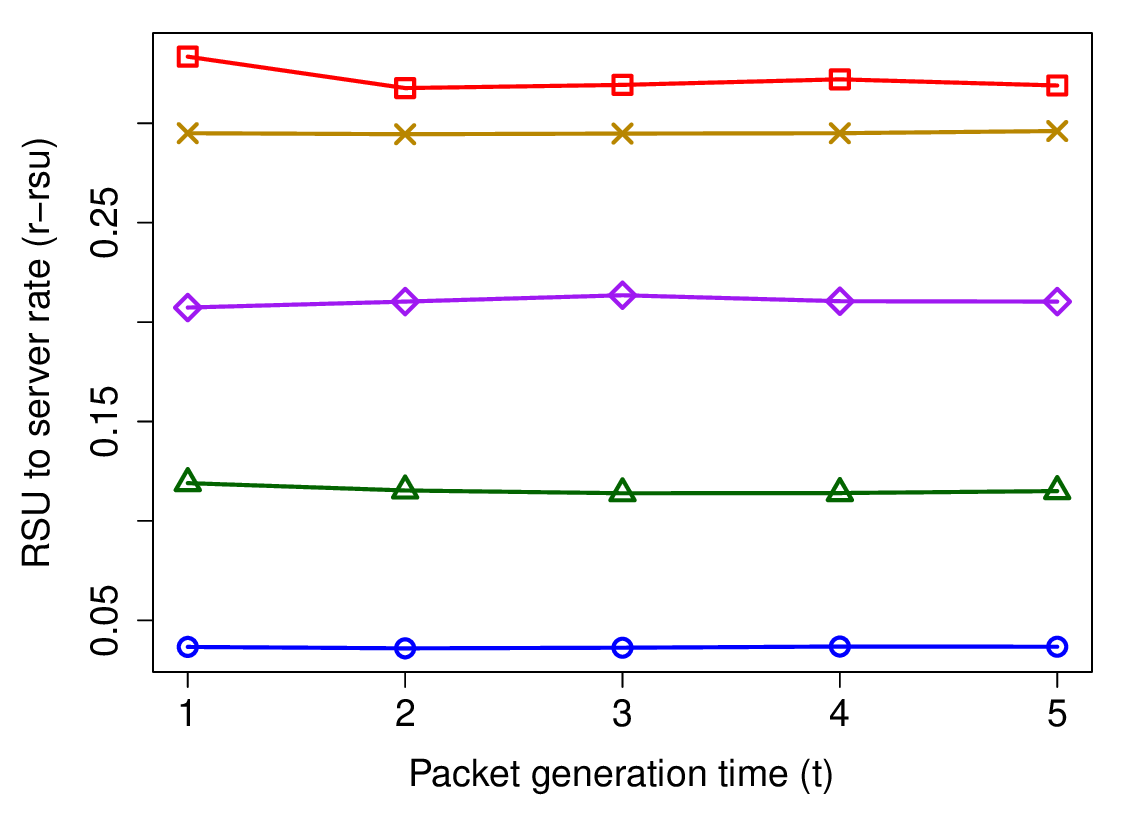}\label{fig:generation_time_3}}
    \caption{Impacts of the packet generation interval ($\lambda_d$). $\delta$ is fixed to 5 time slots. \label{fig:generation_time}} 
\end{figure*}
\begin{figure*}
    \centering
    \includegraphics[width=0.5\textwidth]{figs/journal-update/Legend_new.pdf}\\
    \subfigure[4G communication ratio]{\includegraphics[width=0.3\textwidth]{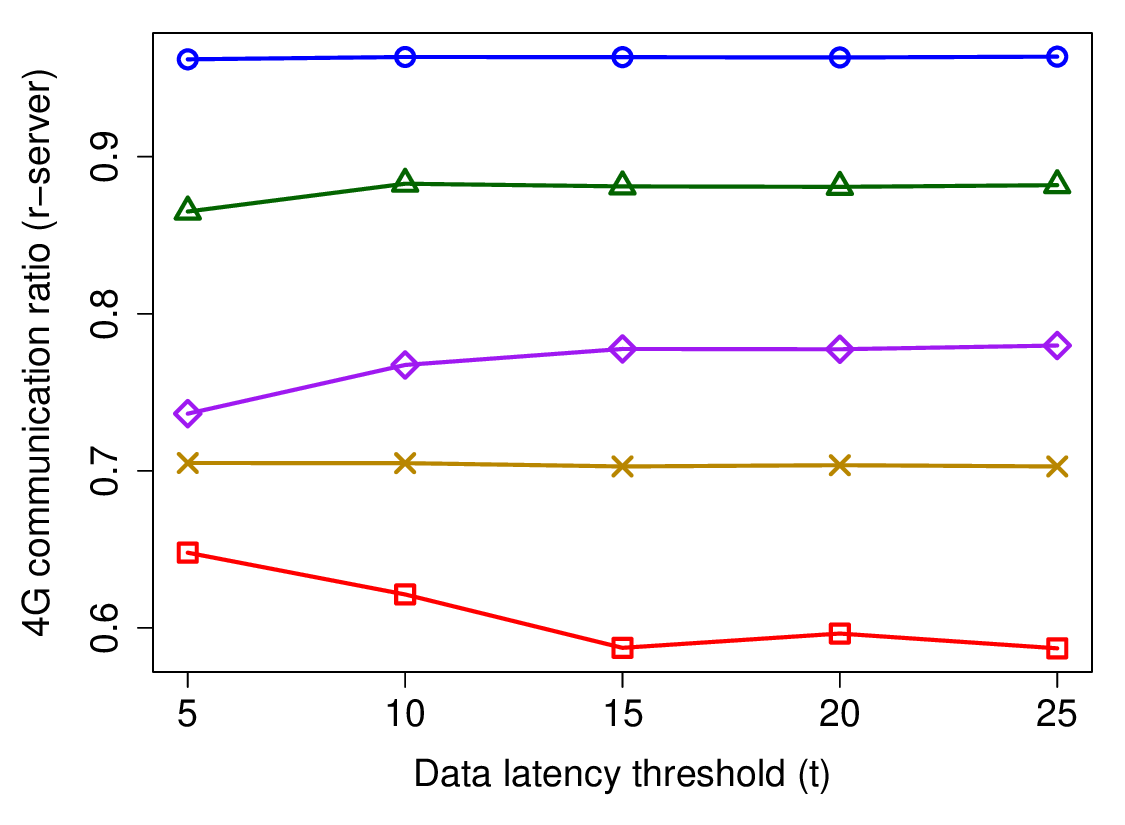} \label{fig:change_delay_1}} 
    \hfill
    \subfigure[Rate of $\delta$-delayed packets]{\includegraphics[width=0.3\textwidth]{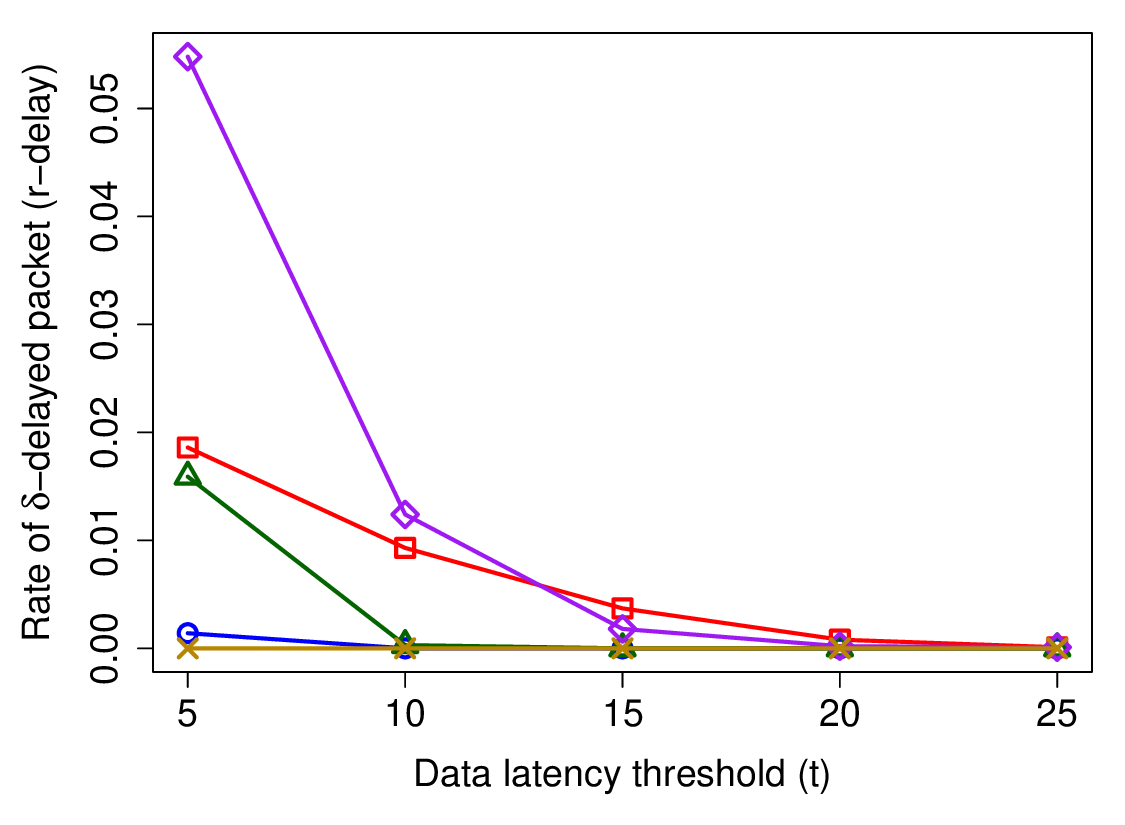}
    \label{fig:change_delay_2}}
    \hfill
    \subfigure[RSU to server rate]{\includegraphics[width=0.3\textwidth]{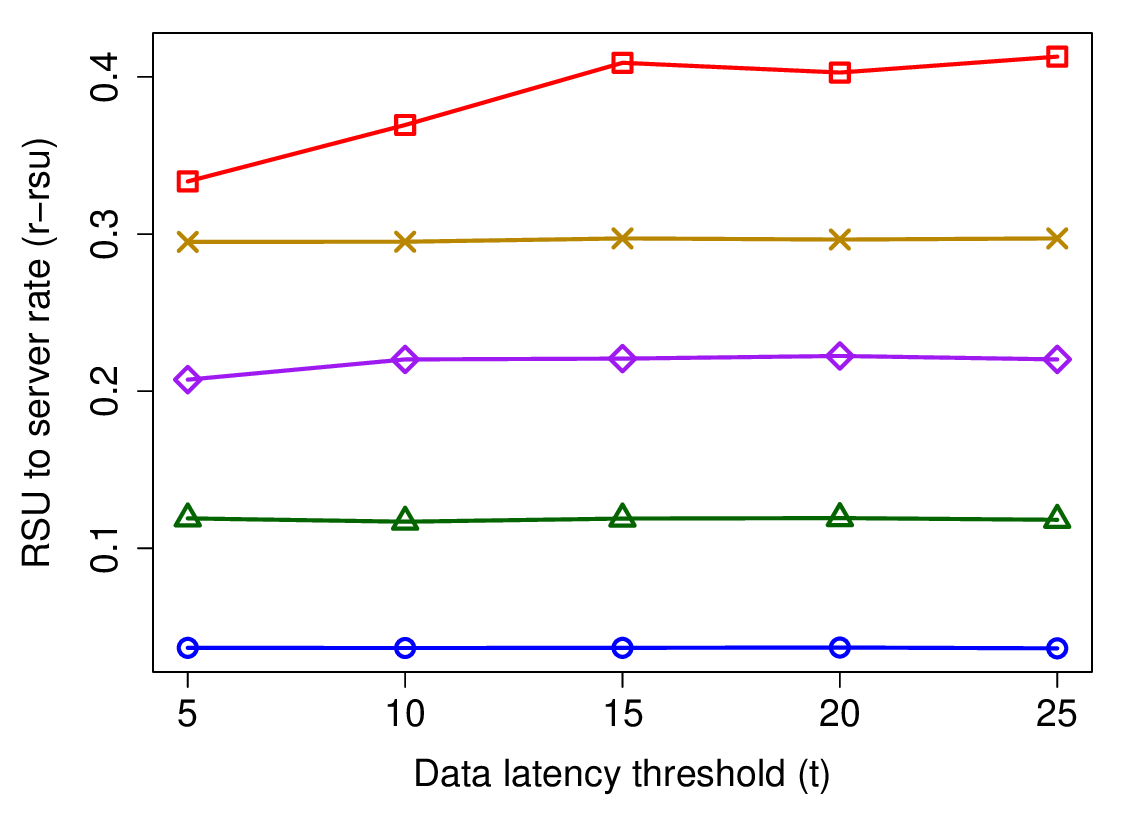} \label{fig:change_delay_3}}
    \caption{Impact of the data latency threshold ($\delta$). $\lambda_d$ is fixed to 1 time slot.}
    \label{fig:change_delay}
\end{figure*}

In summary, the results imply that the performance/cost of the baseline FP approach is sensitive to the environment/network due to its coarse fixed configuration. Thus, it requires effort to manually figure out the best configuration when implementing this method in the real world (for a given specific environment). 
The greedy strategy focuses too much on the rate of news stories without balancing the cost of communication.
By contrast, our proposed method learns the environment/network information to make the decision flexibly so that it can avoid both the packet-dropped issue and \textit{offload-missed} issue while using a smaller communication cost.

\subsection{Discussion}
\label{sec:discussion}

In the following, we investigate the impacts of the packet generation interval, the data latency threshold, the packet size, and the contact rate on our proposed method. 
\subsubsection{Impacts of the Packet Generation Interval $\lambda_d$}
In this evaluation, we estimate the impact of the packet generation interval, i.e., the frequency the devices collect the sensory data, on our proposed method.
We set the data latency threshold to $5$ time slots while changing the packet generation interval from $1 \to 5$ time slots. 
The result in Fig.~\ref{fig:generation_time} shows that the relative communication cost between our proposed method and the baselines does not change. There's a trivial impact of packet generation rate on all the strategies in communication cost. Our proposed method always uses least 4G communication with a small rate of the delayed packet. Although the greedy and FP3 could achieve smaller delayed rates than ours, these two strategies scarify the communication cost. Specifically, FP3 sends 96\% of packets via the 4G network, which is $1.5\times$ higher than that of our proposed method.
When the packet generation interval is slow enough, instead of being dropped, a packet will be stored in a queue and, thus,  the rate of delayed packets is increased. In general, this trend also appears when the relative packet size over the capacity of the local queue is too big (that leads to a smaller number of the packet can be stored in the queue until the queue is full).
Interestingly, the packet generation interval does not affect the packet delayed rate of FP1, FP2 and greedy strategies where the possibility of keeping a packet in the edge devices are not too high, i.e., $P_{rsu} + P_{sensor} = 0.2$ as in FP1 strategy. In contrast, because FP3 and our proposed method try to avoid to use the 4G communication channel as much as possible, the packet generation time has much higher impact on those strategies. For example, to maintain $\approx 63\%$ of 4G communication rate when the packet generation interval changed, our proposed method could not avoid the delayed packet, i.e., around $2-4\%$.
\subsubsection{Impacts of the Data Latency Threshold $\delta$}
In this experiment, we explore the relationship of our proposed method's performance with the data latency threshold,  e.g., the maximum latency required by an application. We fix the packet generation interval $\lambda_d = 1$ while changing the latency threshold $\delta$ from $5 \to 25$ time slots,
Fig.~\ref{fig:change_delay} shows the related results.
When the latency threshold increases the $\delta$-delayed packets rate of our proposed method as well as the baseline strategies decrease. This is an expected result because a packet has a longer time to stay in the local devices. 
Furthermore, let us remind our strategy of the reward function in our Q-Learning method (as shown in the Formula (\ref{eq5})). When the data latency exceeds the threshold, there is a higher possibility of a packet being directly sent to the server using the 4G communication channel by assigning a negative reward.
As the result, when the latency threshold increases, the number of packets that meets such condition and use the 4G communication channel becomes smaller. 
For example, the 4G communication ratio of our proposed method changes from $64.8$\% to $58.7$\% when the latency threshold change from 5 to 25-time slots. This trend does not appear in the greedy and FP strategies, These strategies maintain the 4G communication ratio at around 70\%, 96\%, 88\%, and 77\%. with all the configuration of the data latency threshold.

\subsubsection{Impacts of packet size}
\begin{figure*}
    \centering
    \includegraphics[width=0.5\textwidth]{figs/journal-update/Legend_new.pdf}\\
    \subfigure[4G communication ratio]{\includegraphics[width=0.3\textwidth]{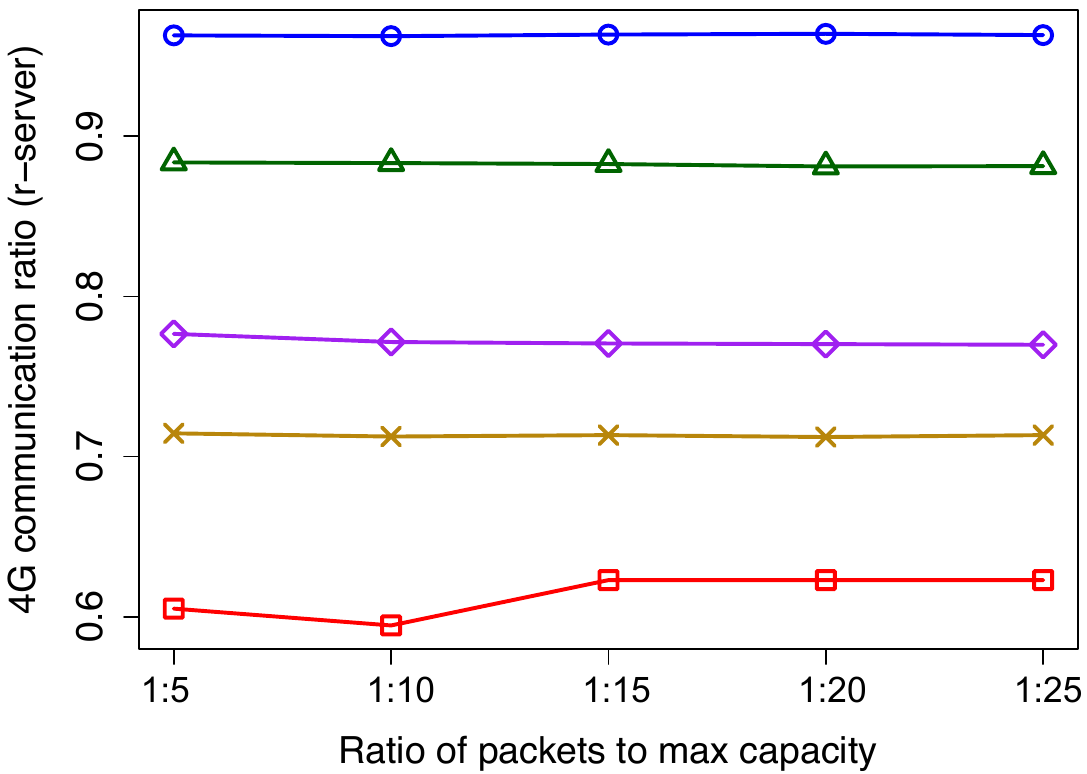}}
    \hfill
    \subfigure[Rate of $\delta$-delayed packets]{\includegraphics[width=0.3\textwidth]{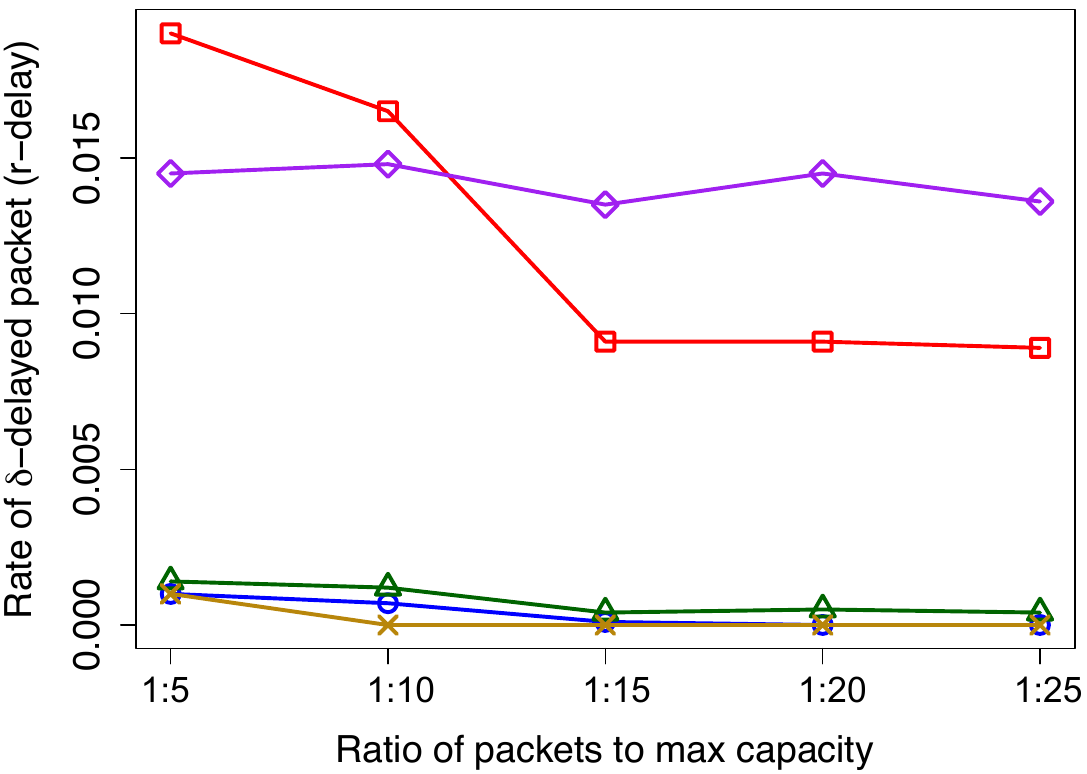}}
    \hfill
    \subfigure[RSU to server rate]{\includegraphics[width=0.3\textwidth]{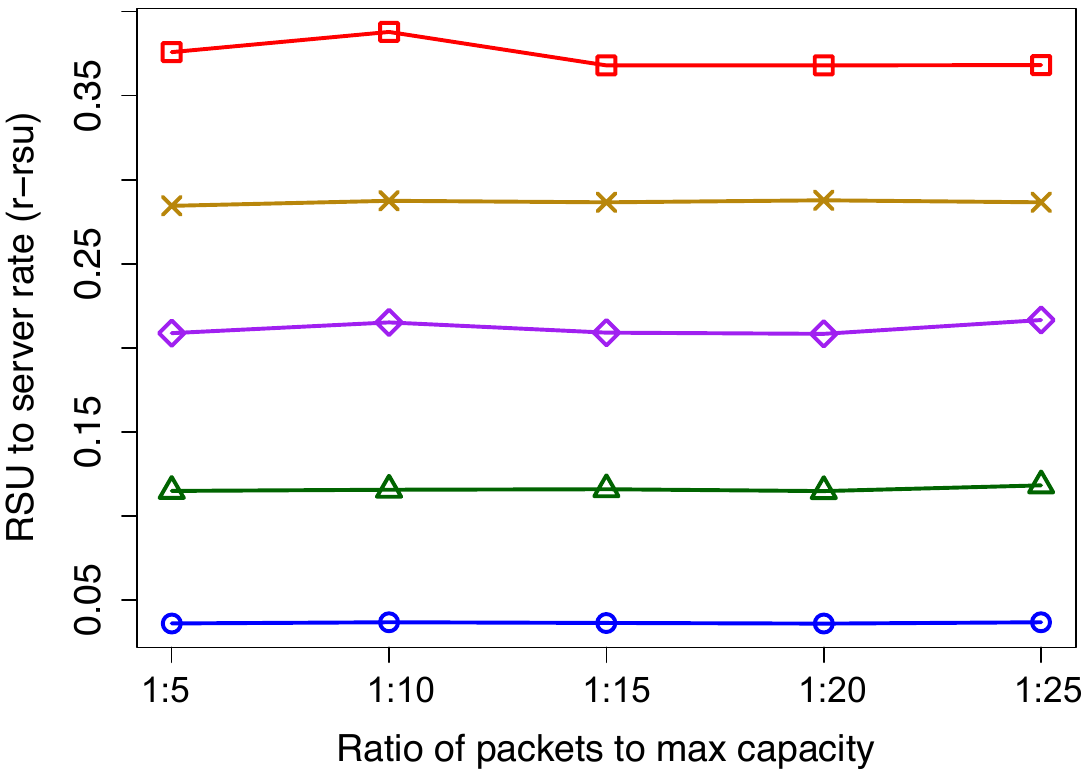}}
    \caption{Impact of the packet size. $\delta$ is fixed to 10 time slots.}
    \label{fig:packet_size}
\end{figure*}

In this section, we investigate the performance of our proposed method when varying the forwarding data's amount, e.g., the packet size from $1 \to 5$ MB. It is worth noting that the packet size affects both the cost of 4G as well as the packet delayed rate because the maximum number of packets stored in the queue decrease when the packet size increases. 
In this evaluation, we fix the queue size of a sensor to 25 MB, i.e., the ratio of packet size to queue size varies from $1:25$ to $1:5$. First of all, we observe that there is no packet dropped, e.g., $r_{drop} = 0$, in all the cases. Second, the result in Fig.~\ref{fig:packet_size} shows that the performance of the greedy and FP baselines is nearly not changed when the packet size is varied. Only our proposed method shows the decrease of ratio $\delta$-delay packet when the packet size becomes smaller because our method dynamically selects the transmission route based on the device’s remaining resource (Formula ~\ref{eq1}).

\subsubsection{Impacts of contact rate}
\begin{table}
    \centering
    \caption{Network contact rate of experiment in Fig.~\ref{fig:contact_rate}} \label{tab:contactRate}
    \begin{tabular}{|c|c|c|c|}
    \hline
         \emph{Sensor transmission range (m)} & \emph{Network contact rate} \\
         \hline
         120 & 0.889\\ 
         100 & 0.868\\ 
         80 & 0.838\\ 
         60 & 0.789\\ 
         40 & 0.714\\ 
         20 & 0.536\\ 
         10 & 0.377\\ 
         \hline
    \end{tabular}
\end{table}

\begin{figure*}
    \centering
   \includegraphics[width=0.5\textwidth]{figs/journal-update/Legend_new.pdf}\\
    \subfigure[4G communication ratio]{\includegraphics[width=0.3\textwidth]{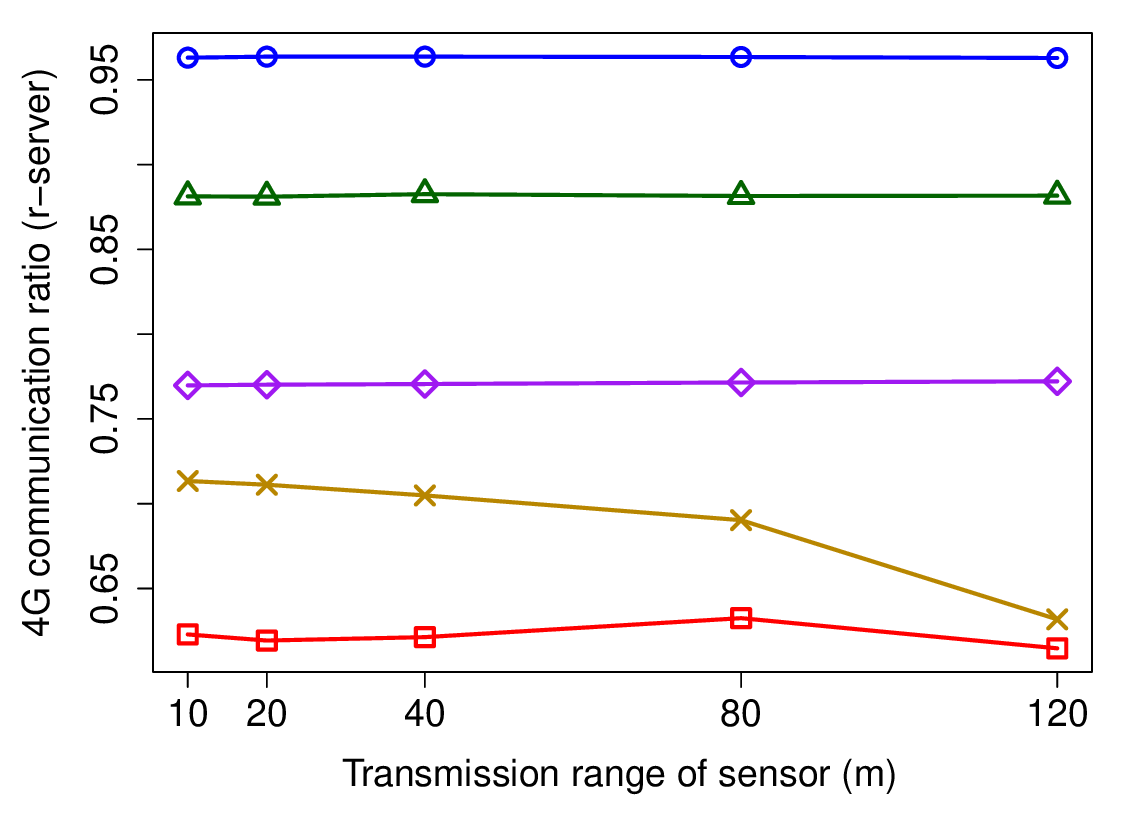}}
    \hfill
    \subfigure[Rate of $\delta$-delayed packets]{\includegraphics[width=0.3\textwidth]{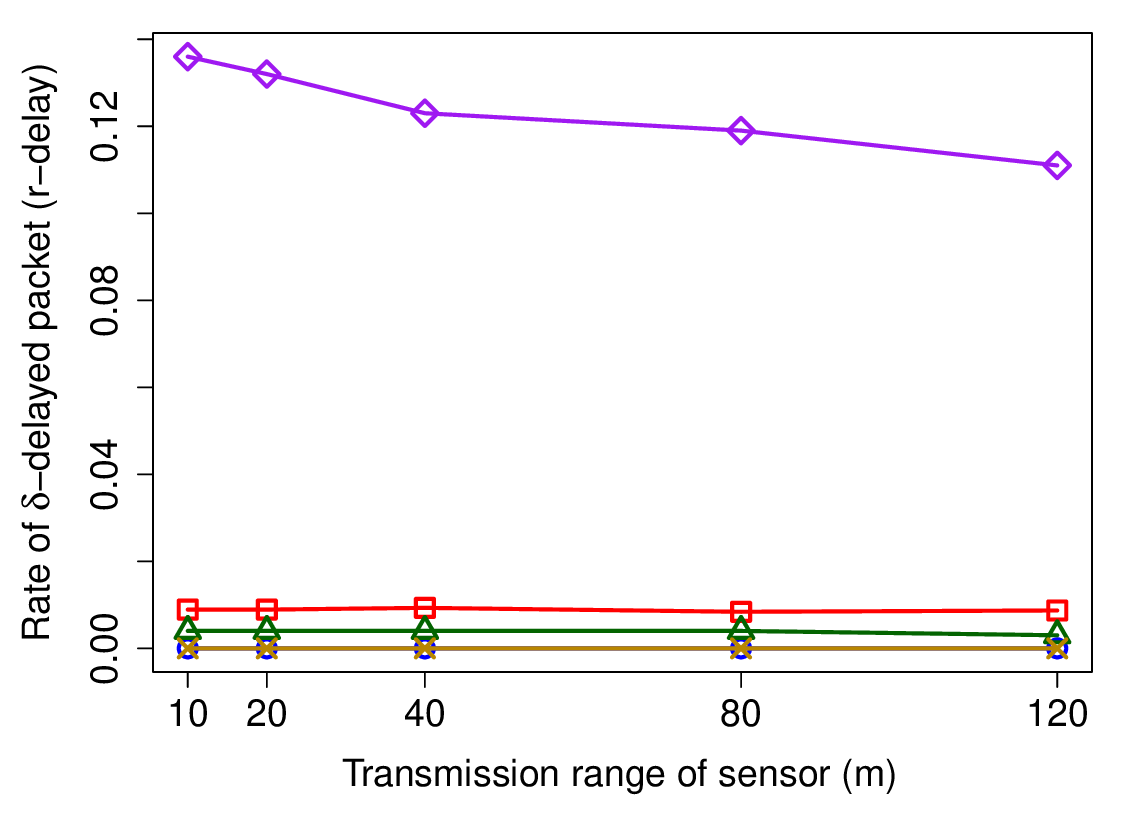}}
    \hfill
    \subfigure[RSU to server rate]{\includegraphics[width=0.3\textwidth]{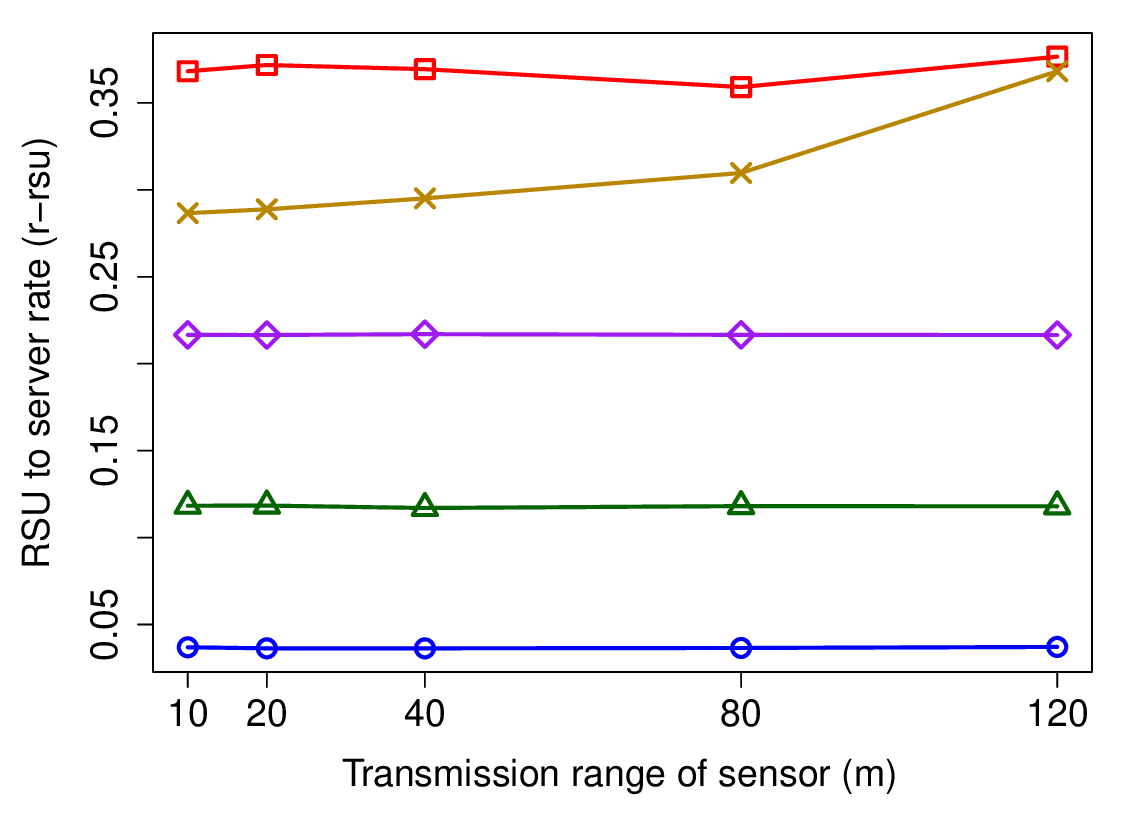}}
    \caption{Impact of the contact rate. $\delta$ is fixed to 10 time slots.}
    \label{fig:contact_rate}
\end{figure*}

In this section, we investigate the impacts of contact rate on the performance of the algorithms. 
We begin by explicitly defining contact rate mathematically.
We refer to each device's contact time as the interval at which it enters the communication area of other devices.
The contact rate of a device is then calculated by dividing its contact time by its entire time on the road.
Furthermore, the network's average contact rate is defined as the average contact rate across all devices.
In this section, we vary the contact rate by changing the transmission range of the devices from $10 \to 120$ meter as shown in Table~\ref{tab:contactRate}. The higher the sensor transmission range, the higher the network contact rate. 
We then observe the change in the 4G communication ratio, delayed packet ratio, and the packet dropped rate.
The results are presented in Fig~\ref{fig:contact_rate}. 
We observe that there is no packet dropped in most of the cases except the FP3, e.g., around 0.01\%. We also confirm that there is no clear correlation between the 4G communication ratio of the FP strategies and the transmission range of devices.
However, when the transmission range of devices increases, the 4G communication ratio of the greedy strategy decreases significantly, e.g., from 71\% to 63\%. In all the configurations, our proposed method always achieves the lowest 4G communication cost. This result implies the stability of the proposed method.

\section{Conclusion} \label{sec:conclusion_and_future_work}
This study focused on real-time vehicular mobile crowdsensing systems which rely on devices mounted on vehicles. 
The devices continuously collect and transmit relevant data to the server through 4G or Wi-Fi communication channels.
We proposed an opportunistic communication algorithm that minimizes the 4G communication cost while guaranteeing the data latency is under a predefined threshold.
We leveraged the Q-learning to make the offloading decision. Besides, the Fuzzy logic is utilized to optimize the reward function of the Q-learning. 
The experiment results indicated that the proposed method could reduce 30-40\% of the 4G communication cost while guaranteeing that 99\% of packets had a latency less than the necessary threshold.


\section*{Acknowledgement}
This work was funded by Vingroup Joint Stock Company (Vingroup JSC), Vingroup, and supported by Vingroup Innovation Foundation (VINIF) under project code VINIF.2021.DA00128, partially funded by Ministry of Education and Training of Vietnam under grant number B2020-BKA-13, and partially funded by Hanoi University of Science and Technology under grant number T2021-PC-019. 

\bibliographystyle{unsrt}
\bibliography{ref}

\begin{thebibliography}{10}

\bibitem{3185504}
Jinwei Liu, Haiying Shen, Husnu~S. Narman, Wingyan Chung, and Zongfang Lin.
\newblock A survey of mobile crowdsensing techniques: A critical component for
  the internet of things.
\newblock {\em ACM Trans. Cyber-Phys. Syst.}, 2(3), June 2018.

\bibitem{7517783}
Jim Cherian, Jun Luo, Hongliang Guo, Shen-Shyang Ho, and Richard Wisbrun.
\newblock Parkgauge: Gauging the occupancy of parking garages with crowdsensed
  parking characteristics.
\newblock In {\em 2016 17th IEEE International Conference on Mobile Data
  Management (MDM)}, volume~1, pages 92--101, 2016.

\bibitem{7192632}
Jun Qin, Hongzi Zhu, Yanmin Zhu, Li~Lu, Guangtao Xue, and Minglu Li.
\newblock Post: Exploiting dynamic sociality for mobile advertising in
  vehicular networks.
\newblock {\em IEEE Transactions on Parallel and Distributed Systems},
  27(6):1770--1782, 2016.

\bibitem{6193237}
Suk-Bok Lee, Joon-Sang Park, Mario Gerla, and Songwu Lu.
\newblock Secure incentives for commercial ad dissemination in vehicular
  networks.
\newblock {\em IEEE Transactions on Vehicular Technology}, 61(6):2715--2728,
  2012.

\bibitem{8703108}
Andrea Capponi, Claudio Fiandrino, Burak Kantarci, Luca Foschini, Dzmitry
  Kliazovich, and Pascal Bouvry.
\newblock A survey on mobile crowdsensing systems: Challenges, solutions, and
  opportunities.
\newblock {\em IEEE Communications Surveys Tutorials}, 21(3):2419--2465, 2019.

\bibitem{9260141}
Hanane Lamaazi, Rabeb Mizouni, Shakti Singh, and Hadi Otrok.
\newblock A mobile edge-based crowdsensing framework for heterogeneous iot.
\newblock {\em IEEE Access}, 8:207524--207536, 2020.

\bibitem{9162762}
Luning Liu, Luhan Wang, and Xiangming Wen.
\newblock Joint network selection and traffic allocation in multi-access edge
  computing-based vehicular crowdsensing.
\newblock In {\em IEEE INFOCOM 2020 - IEEE Conference on Computer
  Communications Workshops (INFOCOM WKSHPS)}, pages 1184--1189, 2020.

\bibitem{8662620}
Pan Zhou, Wenbo Chen, Shouling Ji, Hao Jiang, Li~Yu, and Dapeng Wu.
\newblock Privacy-preserving online task allocation in edge-computing-enabled
  massive crowdsensing.
\newblock {\em IEEE Internet of Things Journal}, 6(5):7773--7787, 2019.

\bibitem{Xia2019}
Xingyou Xia, Yan Zhou, Jie Li, and Ruiyun Yu.
\newblock {Quality-Aware Sparse Data Collection in MEC-Enhanced Mobile
  Crowdsensing Systems}.
\newblock {\em IEEE Transactions on Computational Social Systems},
  6(5):1051--1062, 2019.

\bibitem{Zhao2021}
Yinuo Zhao and Chi~Harold Liu.
\newblock {Social-Aware Incentive Mechanism for Vehicular Crowdsensing by Deep
  Reinforcement Learning}.
\newblock {\em IEEE Transactions on Intelligent Transportation Systems},
  22(4):2314--2325, 2021.

\bibitem{8924663}
Dimitri Belli, Stefano Chessa, Luca Foschini, and Michele Girolami.
\newblock A probabilistic model for the deployment of human-enabled edge
  computing in massive sensing scenarios.
\newblock {\em IEEE Internet of Things Journal}, 7(3):2421--2431, 2020.

\bibitem{9501007}
Yanli Qi, Yiqing Zhou, Zhengang Pan, Ling Liu, and Jinglin Shi.
\newblock Crowd-sensing assisted vehicular distributed computing for hd map
  update.
\newblock In {\em ICC 2021 - IEEE International Conference on Communications},
  pages 1--6, 2021.

\bibitem{9527332}
Chenghao Xu and Wei Song.
\newblock Efficient data uploading for mobile crowdsensing via team
  collaborating and matching.
\newblock {\em IEEE Transactions on Green Communications and Networking}, pages
  1--1, 2021.

\bibitem{8648047}
Piergiorgio Vitello, Andrea Capponi, Claudio Fiandrino, Paolo Giaccone, Dzmitry
  Kliazovich, Ulrich Sorger, and Pascal Bouvry.
\newblock Collaborative data delivery for smart city-oriented mobile
  crowdsensing systems.
\newblock In {\em 2018 IEEE Global Communications Conference (GLOBECOM)}, pages
  1--6, 2018.

\bibitem{9013325}
Wei Gong, Xiaoyao Huang, Guanglun Huang, Baoxian Zhang, and Cheng Li.
\newblock Data offloading for mobile crowdsensing in opportunistic social
  networks.
\newblock In {\em 2019 IEEE Global Communications Conference (GLOBECOM)}, pages
  1--6, 2019.

\bibitem{7553459}
Ke~Zhang, Yuming Mao, Supeng Leng, Quanxin Zhao, Longjiang Li, Xin Peng,
  Li~Pan, Sabita Maharjan, and Yan Zhang.
\newblock Energy-efficient offloading for mobile edge computing in 5g
  heterogeneous networks.
\newblock {\em IEEE Access}, 4:5896--5907, 2016.

\bibitem{platoon_1}
Xiayan Fan, Taiping Cui, Chunyan Cao, Qianbin Chen, and Kyung~Sup Kwak.
\newblock Minimum-cost offloading for collaborative task execution of
  mec-assisted platooning.
\newblock {\em Sensors}, 19(4), 2019.

\bibitem{platoon_3}
Taiping Cui, Yuyu Hu, Bin Shen, and Qianbin Chen.
\newblock Task offloading based on lyapunov optimization for mec-assisted
  vehicular platooning networks.
\newblock {\em Sensors}, 19(22), 2019.

\bibitem{platoon_2}
Hansong Wang, Xi~Li, Hong Ji, and Heli Zhang.
\newblock Federated offloading scheme to minimize latency in mec-enabled
  vehicular networks.
\newblock In {\em 2018 IEEE Globecom Workshops (GC Wkshps)}, pages 1--6, 2018.

\bibitem{2_tier_1}
Jing Zhang, Weiwei Xia, Feng Yan, and Lianfeng Shen.
\newblock Joint computation offloading and urllc resource allocation for
  collaborative mec assisted cellular-v2x networks.
\newblock {\em IEEE Access}, 8:24914--24926, 2020.

\bibitem{3-tier_1}
Junhui Zhao, Qiuping Li, Yi~Gong, and Ke~Zhang.
\newblock Computation offloading and resource allocation for cloud assisted
  mobile edge computing in vehicular networks.
\newblock {\em IEEE Transactions on Vehicular Technology}, 68(8):7944--7956,
  2019.

\bibitem{8402110}
Ying-Dar Lin, Yuan-Cheng Lai, Jian-Xun Huang, and Hsu-Tung Chien.
\newblock Three-tier capacity and traffic allocation for core, edges, and
  devices for mobile edge computing.
\newblock {\em IEEE Transactions on Network and Service Management},
  15(3):923--933, 2018.

\bibitem{Khiem_Le_MEC}
Phi~Le Nguyen, Ren-Hung Hwang, Pham~Minh Khiem, Kien Nguyen, and Ying-Dar Lin.
\newblock Modeling and minimizing latency in three-tier v2x networks.
\newblock In {\em 2020 IEEE Global Communications Conference}, pages 1--6,
  2020.

\bibitem{9679398}
Trung~Thanh Nguyen, Truong Thao~Nguyen, Tuan~Anh Nguyen~Dinh, Thanh-Hung
  Nguyen, and Phi~Le Nguyen.
\newblock Q-learning-based opportunistic communication for real-time mobile air
  quality monitoring systems.
\newblock In {\em 2021 IEEE International Performance, Computing, and
  Communications Conference (IPCCC)}, pages 1--7, 2021.

\bibitem{zadeh1988fuzzy}
Lotfi~A Zadeh.
\newblock Fuzzy logic.
\newblock {\em Computer}, 21(4):83--93, 1988.

\bibitem{simulator}
Opportunistic communication simulator.
\newblock
  https://github.com/AIoT-Lab-BKAI/Fuzzy-Q-learning-based-Opportunistic-Communication,
  04 2022.

\bibitem{jetcheva2003design}
Jetcheva, Hu, PalChaudhuri, Saha, and Johnson.
\newblock Design and evaluation of a metropolitan area multitier wireless ad
  hoc network architecture.
\newblock {\em 2003 Proceedings Fifth IEEE Workshop on Mobile Computing Systems
  and Applications}, pages 32--43, 2003.

\end{thebibliography}


\begin{IEEEbiography}[{\includegraphics[width=1in,height=1.25in,clip,keepaspectratio]{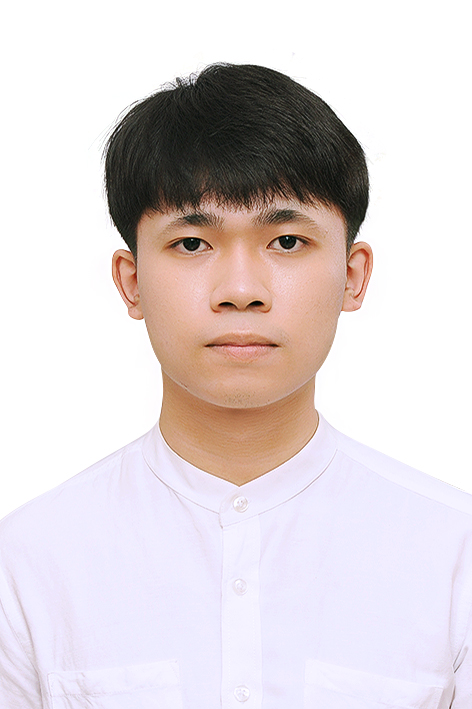}}]{Trung Thanh Nguyen} is a final-year student at the School of Information and Communication Technology at Hanoi University of Science and Technology. He is also a research assistant at the Intelligent Communication Networks laboratory, an International research center for Artificial Intelligence (BK.AI). His research is related to optimization, reinforcement learning, and IoT networks.
\end{IEEEbiography}

\begin{IEEEbiography}[{\includegraphics[width=1in,height=1.25in,clip,keepaspectratio]{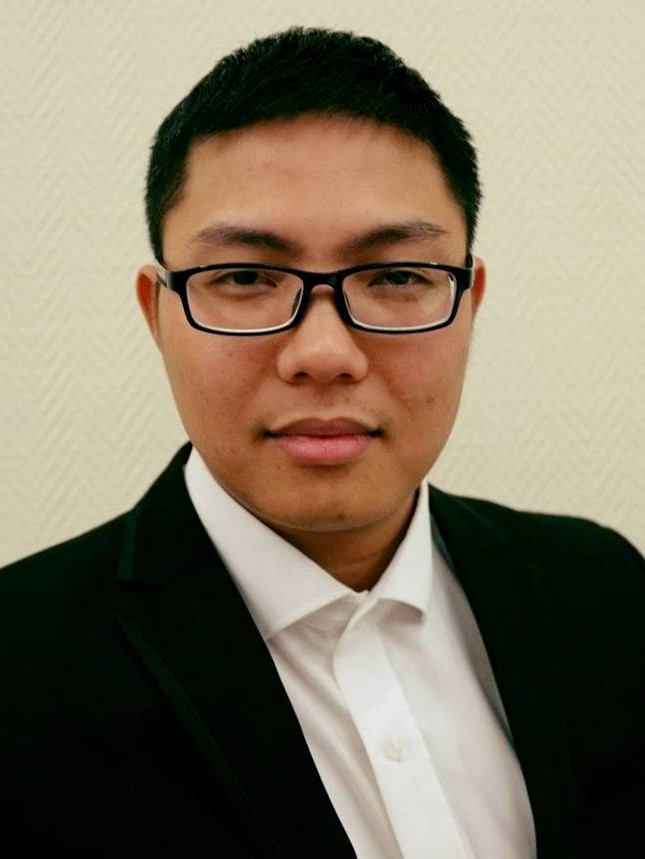}}]{Truong Thao Nguyen}
Dr. Truong Thao Nguyen received the BE and ME degrees from Hanoi University of Science and Technology, Hanoi, Vietnam, in 2011 and 2014, respectively. He received the Ph.D. in Informatics from the Graduate University for Advanced Studies, Japan in 2018. 
 He is currently working at 
 Digital Architecture Research Center, at National Institute of Advanced Industrial Science and Technology (AIST), where he focuses on the topics of High Performance Computing system, Distributed Deep Learning and beyond.
\end{IEEEbiography}

\begin{IEEEbiography}[{\includegraphics[width=1in,height=1.25in,clip,keepaspectratio]{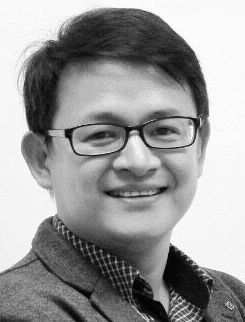}}]{Thanh-Hung Nguyen}
holds a Ph.D. degree in computer science from the University of Grenoble. Currently, he is an assistant professor at School of Information and Communication, Hanoi University of Science and Technology, Vietnam. His research interests are in the Modeling and verification of component-based systems, Network architecture, and Artificial Intelligence.
\end{IEEEbiography}

\begin{IEEEbiography}[{\includegraphics[width=1in,height=1.25in,clip,keepaspectratio]{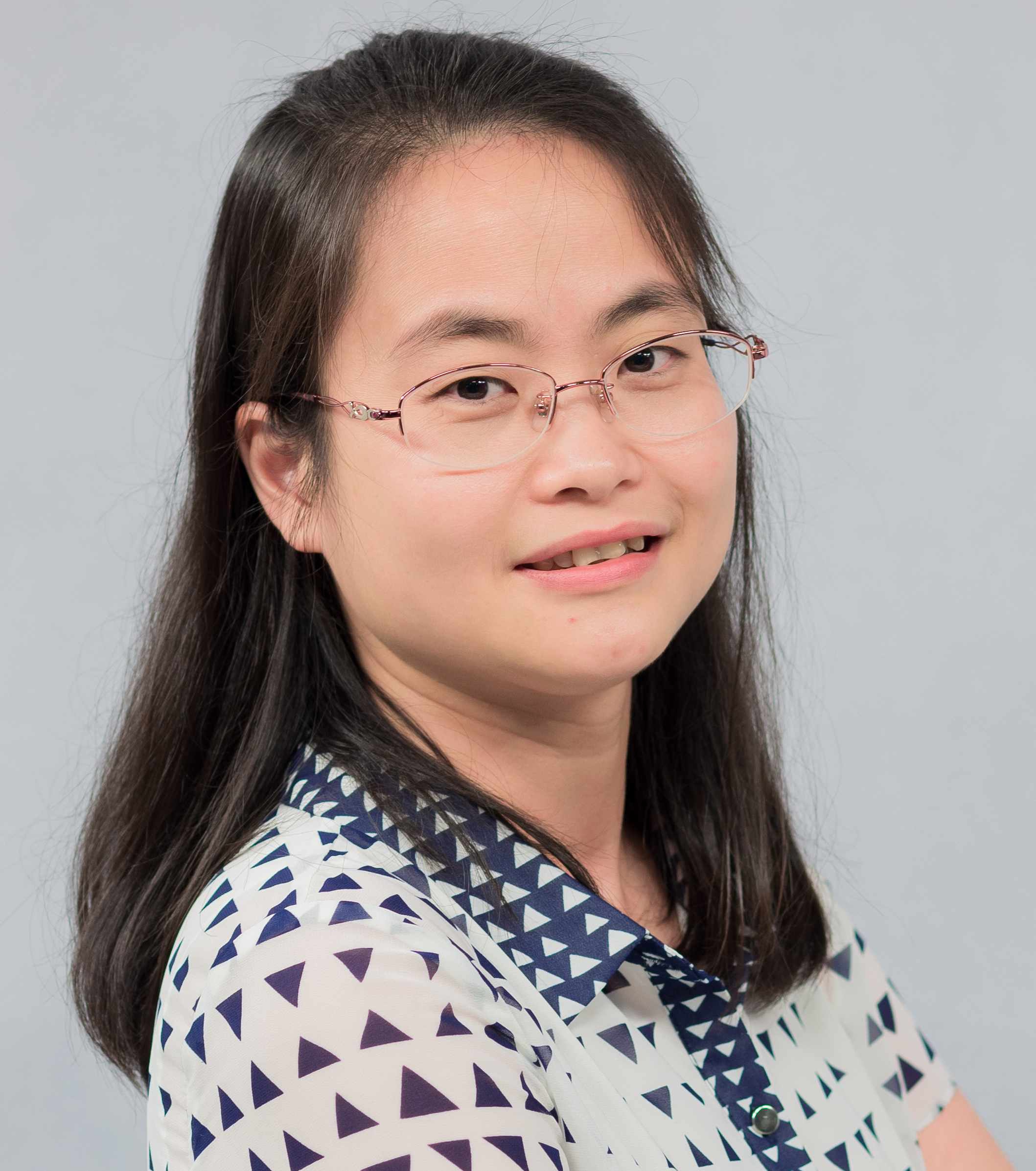}}]{Phi Le Nguyen}
received her B.E. and M.S. degrees from the University of Tokyo in 2007 and 2010, respectively. She received her Ph.D. in Informatics from The Graduate University for Advanced Studies, SOKENDAI, Tokyo, Japan, in 2019. Currently, she is a lecturer at the School of Information and Communication, Hanoi University of Science and Technology (HUST), Vietnam. In addition, she is serving as the managing director at the International research center for Artificial Intelligence (BKAI) and the coordinator of the HEDSPI program, HUST. Her research interests include Network architecture, Optimization, and Artificial Intelligence.
\end{IEEEbiography}

\end{document}